\documentstyle[12pt]{article}

\begin{document}

\baselineskip=16pt plus 1pt minus 1pt

\begin{center}{\large \bf 
E(5) and X(5) critical point symmetries obtained from Davidson potentials  
through a variational procedure } 

\bigskip

{Dennis Bonatsos$^{\#}$\footnote{e-mail: bonat@inp.demokritos.gr},
D. Lenis$^{\#}$\footnote{e-mail: lenis@inp.demokritos.gr}, 
N. Minkov$^\dagger$\footnote{e-mail: nminkov@inrne.bas.bg},
D. Petrellis$^{\#}$,  
P. P. Raychev$^\dagger$\footnote{e-mail: raychev@phys.uni-sofia.bg, 
raychev@inrne.bas.bg}, \break
P. A. Terziev$^\dagger$\footnote{e-mail: terziev@inrne.bas.bg} }
\bigskip

{$^{\#}$ Institute of Nuclear Physics, N.C.S.R.
``Demokritos''}

{GR-15310 Aghia Paraskevi, Attiki, Greece}

{$^\dagger$ Institute for Nuclear Research and Nuclear Energy, Bulgarian
Academy of Sciences }

{72 Tzarigrad Road, BG-1784 Sofia, Bulgaria}

\bigskip

{\bf Abstract}

\end{center} 

Davidson potentials of the form $\beta^2 +\beta_0^4/\beta^2$, when used 
in the E(5) framework (i.e., in the original Bohr Hamiltonian 
after separating variables as in the E(5) model, describing the critical point
of the U(5) to O(6) shape phase transition) bridge the U(5) and O(6) 
symmetries, while they bridge the U(5) and SU(3) symmetries when used in the 
X(5) framework (i.e., in the original Bohr Hamiltonian after separating 
variables as in the X(5) model, corresponding to the critical point 
of the U(5) to SU(3) transition). Using a variational procedure, 
we determine for each value of angular momentum $L$ the value 
of $\beta_0$ at which the rate of change of various physical quantities 
(energy ratios, intraband B(E2) ratios, quadrupole moment ratios) 
has a maximum, the collection of the values of the physical quantity 
formed in this way being a candidate for describing its behavior 
at the relevant critical point. Energy ratios lead to the E(5) and X(5) 
results (whice correspond to an infinite well potential in $\beta$),
while intraband B(E2) ratios and quadrupole moments lead to the 
E(5)-$\beta^4$ and X(5)-$\beta^4$ results, which correspond to the use of 
a $\beta^4$ potential in the relevant framework. 
A new derivation of the Holmberg--Lipas formula for nuclear energy 
spectra is obtained as a by-product. 

\bigskip


\noindent 
Section: Nuclear Structure 

\newpage 

{\bf 1. Introduction} 

The recently introduced E(5) \cite{IacE5} and X(5) \cite{IacX5} models
are supposed to describe shape phase transitions in atomic nuclei, 
the former being related to the transition from U(5) (vibrational) 
to O(6) ($\gamma$-unstable) nuclei, and the latter corresponding to the 
transition from U(5) to SU(3) (rotational) nuclei. In both cases 
the original Bohr collective Hamiltonian \cite{Bohr} is used, with an infinite 
well potential in the collective $\beta$-variable. 
Separation of variables is achieved in the E(5) case by assuming that the 
potential is independent of the collective $\gamma$-variable, while in the 
X(5) case the potential is assumed to be of the form $u(\beta)+u(\gamma)$.     
We are going to refer to these two cases as ``the E(5) framework'' and 
``the X(5) framework'' respectively.
The selection of an infinite well potential in the $\beta$-variable 
in both cases is justified by the fact that the potential is expected 
to be flat around the point at which a shape phase transition occurs.  
Experimental evidence for the occurence of the E(5) and X(5) symmetries in 
some appropriate nuclei is growing (\cite{CZE5,Zamfir}  and 
\cite{CZX5,Kruecken} respectively). 

In the present work we examine if the choice of the infinite well potential 
is the optimum one for the description of shape phase transitions. 
For this purpose, we need one-parameter potentials which can span 
the U(5)-O(6) region in the E(5) framework, as well as the U(5)-SU(3) region
in the X(5) framework. It turns out that 
the exactly soluble \cite{Elliott,Rowe} Davidson potentials \cite{Dav}  
\begin{equation}\label{eq:e1} 
u(\beta) = \beta^2 +{\beta_0^4 \over \beta^2},
\end{equation}
where $\beta_0$ is the position of the minimum of the potential, 
do possess this property. 

Taking into account the fact that various 
physical quantities should change most rapidly at the point of the 
shape phase transition \cite{Werner}, we locate for each value of the 
angular momentum $L$ the value of $\beta_0$ for which the rate of change 
of the physical quantity is maximized. The collection of 
the values of the physical quantity formed in this way 
should then correspond to the behavior of this physical quantity 
at the critical point. As appropriate physical quantities we have used 
energy ratios within the ground state band, as well as intraband B(E2) ratios 
and quadrupole moment ratios within the ground state band, and within excited 
bands. 

The energy ratios within the ground state band lead to results 
very similar to these provided by the infinite well potential in both the 
E(5) and the X(5) frameworks, thus indicating that the choice of the 
infinite well potential in both cases is the optimum one. 
Intraband B(E2) ratios and quadrupole 
moment ratios lead to results close to the ones provided by the 
E(5)-$\beta^4$ \cite{Arias,E5} and X(5)-$\beta^4$ \cite{X5} models, which use 
a $u(\beta)=\beta^4/2$ potential in the E(5) 
and X(5) framework respectively. Further discussion of these results 
is defered to Section 4.  

The variational procedure used here 
is analogous to the one used in the framework of the Variable Moment of 
Inertia (VMI) model \cite{VMI}, where the energy is minimized with respect 
to the (angular momentum dependent) moment of inertia for each value 
of the angular momentum $L$ separately. 

In the framework of the Interacting Boson Model (IBM) \cite{IA}, there have
been attempts to consider the U(5) to O(6) transition \cite{Arias}, 
as well as the U(5) to SU(3) transition \cite{Draayer} in terms  
of one-parameter schematic Hamiltonians. In the present approach, 
in contrast, Davidson potentials are used directly in the original Bohr 
Hamiltonian, without the intervention af any IBM aproximations. 

On the other hand, exactly soluble models have been constructed 
by using the Coulomb and Kratzer \cite{Kratzer} potentials in the original 
Bohr Hamiltonian in the E(5) \cite{FortE5} and X(5) \cite{FortX5} frameworks, 
while a two-parameter quasi-exactly soluble model \cite{TU,TurbCMP,TurbSP} 
has been constructed by using \cite{Levai} the sextic oscillator with a 
centrifugal barrier \cite{Ushv} in the E(5) framework. 
Some of these potentials will be further commented below.  

In Section 2 the E(5) case is considered, while the X(5) case is examined 
in Section 3, in which a new derivation of the Holmberg--Lipas formula 
\cite{Lipas} for nuclear energy spectra, as well as quadrupole moments 
for the X(5) and related models are obtained as by-products. 
Finally, Section 4 contains a discussion of the present results and 
plans for further work. A preliminary version of this work, limited to 
the variational study of the spectra of ground state bands, has been reported 
in Ref. \cite{PLBvar}. 

{\bf 2. Davidson potentials in the E(5) framework } 

{\bf 2.1 Spectra and B(E2) transition rates}

The original Bohr Hamiltonian \cite{Bohr} is
\begin{equation}\label{eq:e2}
H = -{\hbar^2 \over 2B} \left[ {1\over \beta^4} {\partial \over \partial 
\beta} \beta^4 {\partial \over \partial \beta} + {1\over \beta^2 \sin 
3\gamma} {\partial \over \partial \gamma} \sin 3 \gamma {\partial \over 
\partial \gamma} - {1\over 4 \beta^2} \sum_{k=1,2,3} {Q_k^2 \over \sin^2 
\left(\gamma - {2\over 3} \pi k\right) } \right] +V(\beta,\gamma),
\end{equation}
where $\beta$ and $\gamma$ are the usual collective coordinates describing the 
shape of the nuclear surface,
$Q_k$ ($k=1$, 2, 3) are the components of angular momentum, and $B$ is the 
mass parameter. 

Assuming that the potential depends only on the variable $\beta$, 
i.e. $V(\beta,\gamma) = U(\beta)$, one can proceed to separation of variables 
in the standard way \cite{Bohr,WJ1956}, using the wave function 
$ \Psi(\beta,\gamma, \theta_i) = f(\beta) \Phi(\gamma, \theta_i)$,
where $\theta_i$ $(i=1,2,3)$ are the Euler angles describing the orientation 
of the deformed nucleus in space. 

In the equation involving the angles, the eigenvalues of the second order 
Casimir operator of SO(5) occur, having the form 
 $\Lambda = \tau(\tau+3)$, where $\tau=0$, 1, 2, \dots is the quantum 
number characterizing the irreducible representations (irreps) of SO(5), 
called the ``seniority'' \cite{Rakavy}. This equation has been solved 
by B\`es \cite{Bes}.   

The ``radial'' equation can be simplified by introducing \cite{IacE5} 
reduced energies $\epsilon = {2B\over \hbar^2} E$ and reduced potentials 
$u= {2B \over \hbar^2} U$, leading to 
\begin{equation} \label{eq:e3} 
\left[ -{1\over \beta^4} {\partial \over \partial \beta} \beta^4 
{\partial \over \partial \beta} + {\tau(\tau+3) \over \beta^2}+ u(\beta) 
\right] f(\beta) = \epsilon f(\beta). 
\end{equation} 

When plugging the Davidson potentials of Eq. (\ref{eq:e1}) in the above 
equation, the $\beta_0^4 /\beta^2$ term is combined with the 
$\tau(\tau+3)/\beta^2$ term appearing there and the equation is solved 
exactly \cite{Elliott,Rowe}, the eigenfunctions 
being Laguerre polynomials of the form 
\begin{equation}\label{eq:e4} 
F^\tau_n(\beta) = \left[{ 2 n! \over \Gamma 
\left( n+p+{5\over 2}\right)}\right]^{1/2} \beta^p L_n^{p+{3\over 2}}(\beta^2)
e^{-\beta^2/2},   
\end{equation}
where $\Gamma(n)$ stands for the $\Gamma$-function, while 
$p$ is determined by \cite{Elliott}
\begin{equation}\label{eq:e5} 
p(p+3) =\tau(\tau+3) + \beta_0^4,
\end{equation} 
leading to 
\begin{equation}\label{eq:e6} 
p= -{3\over 2} + \left[ \left(\tau+{3\over 2}\right)^2 +\beta_0^4\right]^{1/2}.
\end{equation}
The energy eigenvalues are then \cite{Elliott,Rowe}
(in $\hbar \omega=1$ units) 
\begin{equation}\label{eq:e7} 
E_{n,\tau} = 2n+p+{5\over 2} = 
2n+1+ \left[ \left( \tau+{3\over 2} \right)^2 +\beta_0^4
\right]^{1/2} . 
\end{equation}
For $\beta_0=0$ the original solution of Bohr \cite{Bohr,Dussel}, which 
corresponds to a 5-dimensional (5-D) harmonic oscillator characterized by 
the symmetry U(5) $\supset$ SO(5) $\supset$ SO(3) $\supset$ SO(2) 
\cite{CM870}, is obtained. 
The values of angular momentum $L$ contained in each irrep of SO(5) 
(i.e. for each value of $\tau$) are given by the algorithm \cite{IA} 
$\tau=3\nu_\Delta +\lambda$, where $\nu_\Delta=0$, 1, \dots is the missing 
quantum number in the reduction SO(5) $\supset$ SO(3),  and 
$L=\lambda, \lambda+1, \ldots, 2\lambda-2, 2\lambda$ (with $2\lambda-1$ 
missing). 

The levels of the ground state band are characterized by $L=2\tau$ and $n=0$.
Then the energy levels of the ground state band are given by 
\begin{equation}\label{eq:e8} 
E_{0,L} =1+{1\over 2} \left[ (L+3)^2 +4\beta_0^4\right]^{1/2},
\end{equation}
while the excitation energies of the levels of the ground state band 
relative to the ground state are 
\begin{equation}\label{eq:e9} 
E_{0,L,exc}= E_{0,L}-E_{0,0} ={1\over 2} \left( \left[(L+3)^2
+4\beta_0^4\right]^{1/2} -\left[ 9+4\beta_0^4\right]^{1/2} \right).
\end{equation}

For $u(\beta)$ being a 5-D infinite well 
\begin{equation}\label{eq:e10}
  u(\beta) = \left\{ \begin{array}{ll} 0 & \mbox{if $\beta \leq \beta_W$} \\
\infty  & \mbox{for $\beta > \beta_W$} \end{array} \right.
\end{equation} 
one obtains the E(5) model of Iachello \cite{IacE5} in which the 
eigenfunctions are 
Bessel functions $J_{\tau+3/2}(z)$ (with $z=\beta k$, $k =\sqrt{\epsilon}$), 
while the spectrum is determined by the zeros of the Bessel functions 
\begin{equation}\label{eq:e11}  
E_{\xi,\tau} = {\hbar^2 \over 2B} k^2_{\xi,\tau}, \qquad 
k_{\xi,\tau} = {x_{\xi,\tau} \over \beta_W}
\end{equation}
where $ x_{\xi,\tau}$ is the  $\xi$-th zero of the Bessel function 
$J_{\tau+3/2}(z)$. 
The spectra of the E(5) and Davidson cases become directly comparable 
by establishing the formal correspondence $n=\xi-1$. 

In what follows the ratios 
\begin{equation}\label{eq:e12} 
R_{n,L}= {E_{n,L} -E_{0,0} \over E_{0,2}-E_{0,0} },
\end{equation}
and 
\begin{equation}\label{eq:e13}
\underline{R}_{n,L}= {E_{n,L}-E_{n,0} \over E_{n,2}-E_{n,0}}
\end{equation}
with the notation $E_{n,L}$, will be used.   
In the former case energies in all bands are measured relative to the 
ground state and normalized to the excitation energy of the $L=2$ state 
of the ground state band (as in Ref. \cite{IacE5}), while in the latter case 
energies in each band are measured relative to the bandhead ($L=0$) of this 
band and normalized to the excitation energy of the $L=2$ state of this band.
From Eq. (\ref{eq:e7}) it is clear that Eq. (\ref{eq:e13}) yields identical 
results for all bands with $L=2\tau$, irrespectively of $n$.    
For the ground state band ($n=0$) the simplified notation 
\begin{equation}\label{eq:e14} 
R_L \equiv R_{0,L}
\end{equation}
will also be used. 

The quadrupole operator has the form \cite{WJ1956} 
\begin{equation}\label{eq:e15}
T^{(E2)}_\mu= t \alpha_\mu = 
t \beta \left[ {\cal D}^{(2)}_{\mu,0}(\theta_i) \cos\gamma  +{1\over 
\sqrt{2}} ({\cal D} ^{(2)}_{\mu,2} (\theta_i)+ {\cal D}^{(2)}_{\mu,-2} 
(\theta_i) ) \sin\gamma
\right],
\end{equation}
where $t$ is a scale factor and ${\cal D}(\theta_i)$ denote Wigner functions 
of the Euler angles, while the B(E2) transition rates are given by 
$$
B(E2; \varrho_i L_i\to \varrho_f L_f) =
\frac{1}{2L_i+1}\,|\langle \varrho_f L_f||T^{(E2)}||\varrho_i L_i\rangle|^2 
$$
\begin{equation}\label{eq:e16}
= \frac{2L_f+1}{2L_i+1}\, B(E2; \varrho_f L_f\to \varrho_i L_i), 
\end{equation}
where by $\varrho$ quantum numbers other than the angular momentum $L$ are 
denoted. The calculation of B(E2) rates proceeds as in \cite{IacE5,E5}. 

In what follows, the intraband ratios 
\begin{equation}\label{eq:e17}
R_{n,L}^{B(E2)}= {B(E2; (L+2)_n \to L_n) \over B(E2; 2_0\to 0_0)}
\end{equation}
and 
\begin{equation}\label{eq:e18} 
\underline{R}_{n,L}^{B(E2)}= {B(E2; (L+2)_n \to L_n) \over 
B(E2; 2_n\to 0_n)} 
\end{equation}
will be used. In the former case the B(E2) intraband transition rates
of all bands are normalized to the B(E2) transition rate between the 
two lowest states of the ground state band (as in Ref. \cite{IacE5}), 
while in the latter case 
the B(E2) intraband transition rates within each band are normalized to 
the B(E2) transition rate between the two lowest states of this band. 

It should be noted at this point that quadrupole moments in this framework 
vanish, if one is limited to the quadrupole operator of Eq. (\ref{eq:e15}), 
because of a $\Delta \tau=\pm 1$ selection rule. 
This case is reminiscent of the vanishing (to lowest order) of the quadrupole 
moments in the O(6) limit of IBM \cite{IA}. Non-vanishing quadrupole moments 
can be obtained by including the next order terms in the quadrupole operator 
of Eq. (\ref{eq:e15}).  

{\bf 2.2 The U(5) and O(6) limits} 

For $\beta_0=0$ it is clear that the original vibrational model 
of Bohr \cite{Bohr,Dussel} (with $R_4=2$) is obtained, while for large 
$\beta_0$ the O(6) limit of the Interacting Boson Model (IBM) \cite{IA} 
for large boson numbers, which coincides with the $\gamma$-unstable rotator
(with $R_4=2.5$) is approached \cite{Elliott}. The latter fact can be seen 
in Table 1, where the $R_L$ energy ratios within the ground state band 
for two different values of the 
parameter $\beta_0$ are shown, together with the O(6) predictions
for large boson numbers, which correspond to \cite{Casten} 
\begin{equation}\label{eq:e19} 
E(L)=A L(L+6), \qquad R_L= {L(L+6)\over 16}, 
\end{equation} 
with $A$ constant. It is clear that the O(6) limit is approached as 
$\beta_0$ is increased, the agreement being already very good at $\beta_0=8$.

In Table 1 the intraband B(E2) ratios $\underline{R}_{0,L}^{B(E2)}$ (within 
the ground state band, which has $n=0$) and $\underline{R}_{1,L}^{B(E2)}$ 
(within the next band,
which is characterized by $n=1$) are also shown. In the O(6) limit 
(for infinite number of bosons) one has \cite{IA} 
\begin{equation} \label{eq:e20} 
B(E2; L+2 \to L) = a {L+2 \over L+5},
\end{equation} 
where $a$ constant, 
therefore in this case 
\begin{equation}\label{eq:e21}
\underline{R}_{n,L}^{B(E2)} = {5\over 2} {L+2\over L+5} ,
\end{equation}
for all values of $n$. In Table 1 we remark that the $n=0$ and $n=1$ 
results still differ a little at $\beta_0=4$, becoming almost identical 
to the O(6) behavior at $\beta_0=8$. 

The gradual evolution from the U(5) to the O(6) limit, as $\beta_0$ is 
increased, is depicted in Fig. 1, where the energy ratios $R_L$ within 
the ground state band are depicted, and in Fig. 2, where the 
intraband B(E2) ratios $R_{0,L}^{B(E2)}$ (within the ground state band)
and $R_{1,L}^{B(E2)}$ (within the $n=1$ band) are shown. The limiting 
values at the right hand side are in agreement with the values given in 
Table 1 (taking into account the difference between the $R_{n,L}^{B(E2)}$ 
and $\underline{R}_{n,L}^{B(E2)}$ ratios, defined in Eqs. (\ref{eq:e17}) and 
(\ref{eq:e18})) , while at the left hand side the U(5) values, corresponding to
\cite{E5,IA}
\begin{equation}\label{eq:e22}
E(L) = A L, \qquad R_L= {L \over 2}, 
\end{equation}
\begin{equation}\label{eq:e23}
B(E2;(L+2)_0\to L_0) = a (L+2), \qquad R_{0,L}^{B(E2)} = {L+2\over 2}, 
\end{equation}
\begin{equation}\label{eq:e24} 
B(E2; (L+2)_1 \to L_1) = a' {(L+2) (L+7)\over L+5}, 
\qquad \underline{R}_{1,L}^{B(E2)} = {5\over 14} {(L+2) (L+7)\over L+5}, 
\end{equation}
where $A$, $a$, $a'$ constants, are obtained.   

It should be noticed that the O(6) limit of IBM (with large boson numbers)
is also obtained \cite{FortE5}
by using in the E(5) framework the exactly soluble Kratzer potential
\cite{Kratzer}, 
which has the form 
\begin{equation}\label{eq:e24a}
u(\beta)= -2 {\cal D} \left( {\beta_0\over \beta} -{1\over 2} 
{\beta_0^2\over \beta^2}\right) = -{A\over \beta} + {B\over \beta^2},
\end{equation}  
where ${\cal D}$ is the depth of the minimum of the potential, which 
is located at $\beta_0$, while $A=2 \beta_0 {\cal D}$ and 
$B=\beta_0^2{\cal D}$.  
The O(6) limit is obtained for large values 
of ${\cal D}$ or large values of $\beta_0$. In the case of the Kratzer 
potential, however, it is clear that small values of ${\cal D}$ or small
values of $\beta_0$ lead \cite{FortE5} to the Coulomb potential. 

{\bf 2.3 Variational procedure applied to energy ratios} 

It is useful to consider the ratios $R_L$ within the ground state band, 
defined above,
as a function of $\beta_0$. As seen in Fig. 1, where the ratios $R_4$, 
$R_{12}$ and $R_{20}$ are shown, these ratios increase with $\beta_0$, 
the increase 
becoming very steep at some value $\beta_{0,m}(L)$ of $\beta_0$, 
where the first derivative $dR_L\over d\beta_0$ reaches a maximum value, 
while the second derivative $d^2 R_L\over d\beta_0^2$ vanishes. 
Numerical results for $\beta_{0,m}$ are shown in Table 2, together with 
the values of $R_L$ occuring at these points, which are compared to the 
$R_L$ ratios occuring in the ground state band of the E(5) model \cite{IacE5}. 
Very close agreement of the values determined by the procedure described 
above with the E(5) values is observed in Table 2, as well as in Fig. 3(a),
where these ratios are also shown, together with the corresponding ratios of 
the U(5) (Eq. (\ref{eq:e22})) and O(6) (Eq. (\ref{eq:e19})) limits.   
Finally, the potentials obtained for different angular momenta $L$ are 
depicted in Fig. 4. 

The work performed here is reminiscent of a variational procedure. 
Wishing to determine the critical point in the shape phase transition 
from U(5) to O(6), one chooses a potential (the Davidson potential) 
with a free parameter ($\beta_0$), which helps in covering the whole 
range of interest. Indeed, as we have seen in the previous subsection, 
for $\beta_0=0$ the U(5) picture is obtained, while 
large values of $\beta_0$ lead to the O(6) limit.
One then needs a physical quantity which can serve as a ``measure''
of collectivity.  For this purpose one considers 
the ratios $R_L$, encouraged by the fact that these ratios are well-known
indicators of collectivity in nuclear structure \cite{Mallmann}.  
Since at the critical 
point (if any) one expects the collectivity to change very rapidly, one 
looks, for each $R_L$ ratio separately, for the value of the 
parameter at which the change of $R_L$ is maximum. Indeed, the first derivative
of the ratio $R_L$ with respect to the parameter $\beta_0$
exhibits, as seen in the upper panel of Fig. 1, 
a sharp maximum, which is then a good candidate for being the 
critical point for this particular value of the angular momentum $L$. 
The $R_L$ values at the critical points corresponding 
to each value of $L$ form a collection, which should correspond 
to the behaviour of the ground state band of a nucleus at the critical point.
The infinite well potential used in E(5) succeeds in reproducing all 
these ``critical'' $R_L$ ratios in the ground state band for all 
values of the angular momentum $L$, {\it without using any free parameter}.
It is therefore proved that the infinite well potential is indeed 
the optimum choice for describing the ground state bands of nuclei at the 
critical point of the U(5) to O(6) shape phase transition. 

In other words, starting from the Davidson potentials and using a variational 
procedure, according to which the rate of change of the $R_L$ ratios 
as a function of the parameter $\beta_0$ is 
maximized for each value of the angular momentum $L$ separately, one 
forms the collection of critical values of $R_L$ which corresponds 
to the ground state band of the E(5) model, which is supposed 
to describe nuclei at the critical point.  

Variational procedures in which each value of the angular momentum $L$ is 
treated separately are not unheard of in nuclear physics. An example is 
given by the Variable Moment of Inertia (VMI) model \cite{VMI}, in which 
the energy of the nucleus is minimized with respect to the 
(angular momentum dependent) moment 
of inertia for each value of the angular momentum separately. From 
the cubic equation obtained from this condition, the moment of inertia is 
uniquely determined (as a function of angular momentum) 
in each case. The collection of energy levels occuring 
by using in the energy formula the appropriate value of the 
moment of inertia for each value of the angular momentum $L$ 
forms the ground state band of the nucleus. 

$L$-dependent potentials are also not unheard of in nuclear physics. 
They are known to occur in the framework of the optical model potential
\cite{Fiedel,Mack,Muether}, as well as in the case of quasimolecular 
resonances, like $^{12}$C+$^{12}$C \cite{Scheid}. 

Some comparison of the variational procedure used here with the 
standard Ritz variational method used in quantum mechanics 
(\cite{GM}, for example) is in place. 
In the (simplest version of the) Ritz variational method a trial wave function
containing a parameter is chosen and subsequently the energy is minimized 
with respect to this parameter, thus determining the parameter value 
and, after the relevant substitution, the energy value at the minimum. 
In the present case a trial potential containing a parameter is chosen 
and subsequently the rate of change of the 
physical quantity (here the rate of change of the energy ratios) 
is maximized with respect to this parameter, thus determining the 
parameter value and, after the relevant calculation, the value of the 
physical quantity (here the energy ratios) at the maximum (i.e. at the 
critical point). 
The main similarity between the two methods is the use of a 
parameter-dependent trial wave function/trial potential respectively. 
The main difference between the two methods is that in the former 
the relevant physical quantity (the energy) is minimized with respect 
to the parameter, while in the latter the rate of change of the physical 
quantity (the energy ratios) is maximized with respect to the parameter.  

It should be emphasized that the trial potentials to be used 
for the study of the critical region between two different symmetries 
should possess the correct limiting behavior, i.e. they should be able 
to reproduce the two symmetries for special values of the free parameter
(as in the present case the Davidson potentials reproduce the U(5) 
symmetry for $\beta_0=0$ and the O(6) symmetry for large $\beta_0$). 

It is worth mentioning at this point that the consequences of replacing in 
the E(5) framework the infinite well potential by a well of finite depth have 
been studied in detail \cite{Caprio}, the main conclusion being that many key 
features of E(5) remain essentially unchanged, even if the depth of the 
potential is radically changed. This observation implies that the 
E(5) predictions, reassured above through the variational procedure, 
are stable and do not depend sensitively on any parameter like the depth 
of the potential. 

The success of the variational procedure when applied to energy ratios 
within the ground state band encourages its use for excited bands 
as well. Since it is reasonable to treat each band as a separate entity, 
the ratios $\underline{R}_{n,L}$ (defined in Eq. (\ref{eq:e13})) should be 
used for this purpose, which involve 
levels of one band only, in contrast to the ratios $R_{n,L}$ 
(defined in Eq. (\ref{eq:e12})) which, 
 except in the case of the ground state band, use levels from two 
different bands. From Eq. (\ref{eq:e7}) it is clear, however, that 
in the case of the Davidson potential and for bands with $L=2\tau$ the 
$\underline{R}_{n,L}$ ratios will be identical to $R_L$ for all values of $n$. 
This is a special feature of the Davidson potentials, due to the 
their oscillator-like spectrum. This feature is lifted when one considers 
generalized Davidson potentials of the form 
\begin{equation}\label{eq:e25}
u(\beta)= \beta^{2n} + {\beta_0^{4n} \over \beta^{2n}}, \quad 
n=2,3,4, \dots 
\end{equation} 
which will be shortly discussed in Section 4. 

{\bf 2.4 Variational procedure applied to excitation energies} 

It is instructive to apply the variational procedure developed 
in the previous subsection to isolated energy levels instead of energy 
ratios. For this purpose the excitation energies $E_{0,L,exc}$ of the levels 
of the ground state band, given by Eq. (\ref{eq:e9}), will be used. The values 
$\beta_{0,m}$ where the absolute value of the 
first derivative (since the first derivative is negative in this case) 
becomes maximum are reported 
in Table 2, together with the corresponding $E_{0,L,exc}$ values. 
Using the $E_{0,L,exc}$ values obtained in this way, one can calculate 
the relevant $R_L$ ratios of Eq. (\ref{eq:e14}). As seen in Table 2, 
the results obtained in this way are very close to the U(5) results, 
provided by Eq. (\ref{eq:e22}). 

This result is easy to explain: From the experimental data (see Ref. 
\cite{Sakai}, for example) it is known that excitation energies 
within a series of isotopes drop very rapidly in the region of the 
vibrational limit, as one moves away from the pure vibrational behaviour
(see, for example, the chains of the Sm, Gd, Dy isotopes),  
while the changes near the rotational limit as one moves from one isotope 
to the next are minimal (see, for example, the Th, U, Pu isotopes). 
Trying then to identify a series of energy levels corresponding to 
the most rapid changes in the excitation energies, one naturally ends 
up with the vibrational limit. Therefore the application of the variational
procedure to isolated energy levels just demonstrates the effectiveness 
of the method, leading to results physically expected. 

{\bf 2.5 Variational procedure applied to B(E2) ratios} 

The success of the variational procedure when applied to the energy ratios 
$R_L$ also encourages its use for intraband B(E2) ratios. Considering each 
band as a separate entity, it is reasonable to use the ratios 
$\underline{R}_{n,L}^{B(E2)}$, defined in Eq. (\ref{eq:e18}), which involve 
transitions 
within one band (while the ratios $R_{n,L}^{B(E2)}$, defined in 
Eq. (\ref{eq:e17}), involve intraband 
transitions from two different bands, except in the case of the 
ground state band). It should be emphasized that the different choice of the 
denominator in the ratios $R_{n,L}^{B(E2)}$ and $\underline{R}_{n,L}^{B(E2)}$ 
is not a trivial matter of normalization, since one divides in each case 
by a different function of $\beta_0$, being led in this way to different 
results when the variational procedure is applied. 

As we have seen in Fig. 2, the B(E2) ratios go down from their U(5) values 
at $\beta_0=0$ to the O(6) limiting values at large $\beta_0$. Therefore 
in this case we are going to determine the values $\beta_{0,m}$ at which 
the absolute value of the first derivative, 
$| d \underline{R}_{n,L}^{B(E2)} /d \beta_0 |$ 
has a maximum, while the second derivative vanishes. Numerical results 
for the ground state band ($n=0$) and the $n=1$ band are given in Table 2, 
and depicted in Figs. 3(b) and 3(c) respectively, where the U(5) and O(6) 
results, calculated 
from Eqs. (\ref{eq:e23}), (\ref{eq:e24}), and (\ref{eq:e21}), are shown for 
comparison. In addition, the results given by the original E(5) model 
\cite{IacE5}, as well as by the E(5)-$\beta^4$ model \cite{Arias,E5}, 
which uses a $u(\beta)=\beta^4/2$ potential in the E(5) framework 
instead of an infinite well potential, are exhibited. In both bands 
the variational procedure leads to results which are quite close 
to the E(5)-$\beta^4$ case. We defer further discussions of these results 
to Section 4. 

It should be mentioned at this point that the first derivative of the 
$B(E2:L+2\to L)$ values with respect to $\beta_0$ does not exhibit 
a maximum, while the second derivative does not vanish at any value 
other than $\beta_0=0$. Therefore the variational procedure cannot be applied 
to isolated B(E2) values, being applicable to B(E2) ratios only. 

{\bf 3. Davidson potentials in the X(5) framework} 

{\bf 3.1 Spectra, B(E2) transition rates, and quadrupole moments} 

Starting again from the original Bohr Hamiltonian of Eq. (\ref{eq:e2}), 
one seeks solutions of the relevant Schr\"odinger equation having 
the form 
$ \Psi(\beta, \gamma, \theta_i)= \phi_K^L(\beta,\gamma) 
{\cal D}_{M,K}^L(\theta_i)$, 
where $\theta_i$ ($i=1$, 2, 3) are the Euler angles, ${\cal D}(\theta_i)$
denote Wigner functions of them, $L$ are the eigenvalues of angular momentum, 
while $M$ and $K$ are the eigenvalues of the projections of angular 
momentum on the laboratory-fixed $z$-axis and the body-fixed $z'$-axis 
respectively. 

As pointed out in Ref. \cite{IacX5}, in the case in which the potential 
has a minimum around $\gamma =0$ one can write  the last term of Eq. 
(\ref{eq:e2}) in the form 
\begin{equation}\label{eq:e26} 
\sum _{k=1,2,3} {Q_k^2 \over \sin^2 \left( \gamma -{2\pi \over 3} k\right)}
\approx {4\over 3} (Q_1^2+Q_2^2+Q_3^2) +Q_3^2 \left( {1\over \sin^2\gamma}
-{4\over 3}\right).  
\end{equation}
Using this result in the Schr\"odinger equation corresponding to 
the Hamiltonian of Eq. (\ref{eq:e2}), introducing reduced energies 
 $\epsilon = 2B E /\hbar^2$ and reduced potentials $u = 2B V /\hbar^2$,  
and assuming that the reduced potential can be separated into two terms, 
one depending on $\beta$ and the other depending on $\gamma$, i.e. 
$u(\beta, \gamma) = u(\beta) + u(\gamma)$, the Schr\"odinger equation can 
be separated into two equations \cite{IacX5,Bijker}, the ``radial'' one being
\begin{equation} \label{eq:e27}
\left[ -{1\over \beta^4} {\partial \over \partial \beta} \beta^4 
{\partial \over \partial \beta} + {1\over 4 \beta^2} {4\over 3} 
(L(L+1)-K^2) +u(\beta) \right] \xi_L(\beta) =\epsilon_\beta  \xi_L(\beta). 
\end{equation}

When plugging the Davidson potentials of Eq. (\ref{eq:e1}) in this equation, 
the $\beta_0^4 /\beta^2$ term of the potential is combined with the 
$(L(L+1)-K^2)/3\beta^2$ term appearing there and the equation is solved 
exactly, the eigenfunctions being Laguerre polynomials of the form 
\begin{equation}\label{eq:e28} 
F^L_n(\beta) = \left[{ 2 n! \over \Gamma 
\left( n+a+{5\over 2}\right)}\right]^{1/2} \beta^a L_n^{a+{3\over 2}}(\beta^2)
e^{-\beta^2/2}  
\end{equation}
where $a$ is given by
\begin{equation}\label{eq:e29} 
a= -{3\over 2} + \left[ {1\over 3} (L(L+1)-K^2) + {9\over 4} 
+\beta_0^4\right]^{1/2}.
\end{equation}
The energy eigenvalues are then (in $\hbar \omega=1$ units) 
\begin{equation}\label{eq:e30} 
E_{n,L}^{(K)} = 2n+a +{5\over 2} = 
2n+1+ \left[ {1\over 3} (L(L+1)-K^2) + {9\over 4} +\beta_0^4
\right]^{1/2} . 
\end{equation}

The levels of the ground state band are characterized by $n=0$ and $K=0$.
Then the excitation energies relative to the ground state are 
given by 
\begin{equation}\label{eq:e31} 
E_{0,L,exc}^{(0)} = E_{0,L}^{(0)}-E_{0,0}^{(0)} = 
\left[{1\over 3} L(L+1) + {9\over 4} +\beta_0^4\right]^{1/2}
-\left[ {9\over 4} +\beta_0^4\right]^{1/2},
\end{equation}
which can easily be put into the form 
\begin{equation}\label{eq:e32} 
E'_{0,L,exc} = {E_{0,L,exc}^{(0)} \over \left[ {9\over 4} 
+\beta_0^4\right]^{1/2} }= \left[1+ {L(L+1)\over 3 \left( {9\over 4} 
+\beta_0^4 \right) }\right]^{1/2} -1,
\end{equation}
which is the same as the Holmberg--Lipas formula \cite{Lipas} 
\begin{equation}\label{eq:e33}
E_{H}(L) = a_{H} \left( \sqrt{1+b_{H}  L(L+1)} -1 \right), 
\end{equation}
with
\begin{equation}\label{eq:e34}
a_H=1, \qquad b_H = {1\over 3\left( {9\over 4} +\beta_0^4\right)  } . 
\end{equation}

It is worth remarking at this point that the Holmberg--Lipas formula
can be derived \cite{Casten} by assuming that the moment of inertia $I$ in the 
energy expression of the rigid rotator ($E(L)=L(L+1)/2I$) is a function of the 
excitation energy, i.e. $I= \alpha + \beta E(L)$, where $\alpha$ and 
$\beta$ are constants, the latter being proportional to $b_H$ and 
acquiring positive values. It is therefore clear that the Holmberg--Lipas 
formula, as well as the spectrum of the Davidson potentials derived 
in this section, have built-in the concept of the Variable Moment 
of Inertia (VMI) model \cite{VMI}, according to which the moment of inertia 
is an increasing function of the angular momentum. 

For $u(\beta)$ being a 5-D infinite well potential (see Eq. (\ref{eq:e10}))
one obtains the X(5) model of Iachello \cite{IacX5}, 
in which the eigenfunctions are Bessel functions $J_\nu(k_{s,L}\beta)$ with
\cite{IacX5,Bijker}  
\begin{equation}\label{eq:e35}
\nu=\left( {L(L+1)-K^2\over 3}+{9\over 4}\right)^{1/2},
\end{equation}
while the spectrum is determined by the zeros of the Bessel functions, 
the relevant eigenvalues being
\begin{equation}\label{eq:e36}
\epsilon_{\beta; s,L} = (k_{s,L})^2, \qquad 
k_{s,L}=  {x_{s,L} \over \beta_W},
\end{equation}
where $x_{s,L}$ is the $s$-th zero of the Bessel function 
$J_\nu(k_{s,L}\beta)$. The eigenfunctions are 
\begin{equation}\label{eq:e37} 
\xi_{s,L}(\beta) = c_{s,L} \beta^{-3/2} J_\nu(k_{s,L}\beta),
\end{equation}
where $c_{s,L}$ are normalization constants. 

The spectra of the X(5) and Davidson cases become 
directly comparable by establishing the formal correspondence 
$n=s-1$. 

In addition to the energy ratios $R_{n,L}$ and $\underline{R}_{n,L}$, 
defined in Eqs. (\ref{eq:e12}) and (\ref{eq:e13}), which will be used 
for $K=0$ bands, the ratios 
\begin{equation}\label{eq:e37a}
\underline{R'}_{n,L} = {E_{n,L}^{(2)} - E_{n,2}^{(2)} \over E_{n,3}^{(2)} 
-E_{n,2}^{(2)} } , 
\end{equation}
defined within the $K=2$ band, will be used below. 

The quadrupole operator is again given by Eq. (\ref{eq:e15}), while 
the B(E2) transition rates are given by 
\begin{equation}\label{eq:e38} 
 B(E2; L_s\to L'_{s'})= {| \langle L_s || T^{(E2)}  || L'_{s'} \rangle |^2 
\over 2L_s+1}.
\end{equation}
The matrix elements of the quadrupole  operator 
involve an integral over the Euler angles, which is the same as in Ref. 
\cite{IacX5} and is performed by using the properties of the Wigner 
${\cal D}$ functions, of which only ${\cal D}_{\mu,0}^{(2)}$ participates, 
since $\gamma\simeq 0$ in Eq. (\ref{eq:e15}) (as mentioned before Eq. 
(\ref{eq:e26})), as well as an integral over $\beta$. After performing 
the integrations over the angles one is left with 
\begin{equation}\label{eq:e39}
B(E2; L_s \to L'_{s'}) = (L_s 2 L'_{s'} | 000)^2  I^2 _{s,L; s', L'}, 
\end{equation}
where the Clebsch--Gordan coefficient $(L_s 2 L'_{s'}| 000)$ appears, 
which determines the relevant selection rules. 
In the Davidson case  the integral has the form 
\begin{equation}\label{eq:e40}
I_{s,L; s', L'} = \int \beta F^L_n(\beta) F^{L'}_{n'}(\beta) \beta^4 d\beta, 
\end{equation}  
with $n=s-1$ and $n'=s'-1$, which 
involves Laguerre polynomials, as seen from Eq. (\ref{eq:e28}). 

In addition to the intraband B(E2) ratios defined in Eqs. (\ref{eq:e17}) 
and (\ref{eq:e18}), interband B(E2) transition rate ratios 
\begin{equation}\label{eq:e41} 
R^{B(E2)}_{n,L,n',L'} = {B(E2;(L_n \to L'_{n'}) \over B(E2; 2_0\to 0_0)} 
\end{equation}
will be used. 

Quadrupole moments are defined by \cite{IA}
\begin{equation}\label{eq:e42} 
Q_{s,L} = {4\sqrt{\pi} \over 5} (L_s L_s 2| L_s  -L_s  0) \langle L_s ||
T^{(E2)} || L_s\rangle .
\end{equation}

In what follows, the ratios 
\begin{equation}\label{eq:e43}
R^Q_{n,L}  = {Q_{n,L} \over Q_{0,2} },
\end{equation}
and 
\begin{equation}\label{eq:e44} 
\underline{R}^Q_{n,L} = {Q_{n,L} \over Q_{n,2} }
\end{equation} 
will be used. 

{\bf 3.2 The X(5)-$\beta^2$ and SU(3) limits} 

For $\beta_0=0$ the exactly soluble X(5)-$\beta^2$ model
(with $R_4 = 2.646$) is obtained, the details of which can be found in 
Ref. \cite{X5}, while for large $\beta_0$ the SU(3) limit of IBM with 
large boson numbers, which coincides with the rigid rotator 
(with $R_4=3.333$) is obtained. In what follows the occurence of the SU(3) 
limit will be discussed in more detail.

It is clear that the Holmberg--Lipas formula gives rotational spectra 
for small values of $b_H$, at which one can keep only the first 
$L$-dependent term 
in the Taylor expansion of the square root appearing in Eq. (\ref{eq:e33}),
leading to energies proportional to $L(L+1)$. From Eq. (\ref{eq:e34})
it is then clear that rotational spectra are expected for large values 
of $\beta_0$, for which small values of $b_H$ occur.  
This can be seen in Table 3, where the $R_L$ ratios 
occuring for two different values of $\beta_0$ are shown, together 
with the predictions of the SU(3) limit of IBM at large boson numbers, 
which correspond to the rigid rotator with 
\begin{equation}\label{eq:e45}
E(L)=A L(L+1), \qquad R_L= {L(L+1) \over 6}, 
\end{equation} 
where $A$ constant \cite{IA}. 
The agreement to the SU(3) results is quite good already at $\beta_0=8$. 
The same is seen for the ratios $\underline{R'}_{0,L}$, regarding the 
$n=0$, $K=2$ band, also reported in Table 3, which in the rigid 
rotator case correspond to the limiting values 
\begin{equation}\label{eq:e46a}
\underline{R'}_{0,L}= {L(L+1)\over 6}-1. 
\end{equation} 

In Table 3 the intraband B(E2) ratios $\underline{R}_{0,L}^{B(E2)}$ (within 
the ground state band, which has $n=0$) and $\underline{R}_{1,L}^{B(E2)}$ 
(within the next band,
which is characterized by $n=1$) are also shown. In the SU(3) limit 
(for infinite number of bosons) one has for all $K=0$ bands \cite{IA} 
\begin{equation} \label{eq:e46} 
B(E2; L+2 \to L) = a {(L+2)(L+1) \over (2L+3)(2L+5)},
\end{equation} 
where $a$ constant, 
therefore in this case 
\begin{equation}\label{eq:e47}
\underline{R}_{n,L}^{B(E2)} = {15\over 2} {(L+2)(L+1)\over (2L+3)(2L+5)} ,
\end{equation}
for all values of $n$. In Table 3 we remark that the $n=0$ and $n=1$ 
results still differ a little at $\beta_0=4$, becoming almost identical 
to the SU(3) behavior at $\beta_0=8$. 

The gradual evolution from the X(5)-$\beta^2$ to the SU(3) limit, as 
$\beta_0$ is increased, is depicted in Fig. 5, where the energy ratios $R_L$ 
within the ground state band are depicted, and in Fig. 6(a), where the 
intraband B(E2) ratios $R_{0,L}^{B(E2)}$ (within the ground state band)
and $R_{1,L}^{B(E2)}$ (within the $n=1$ band) are shown. The limiting 
values at the right hand side are in agreement with the results given in 
Table 3 (taking into account the difference between the ratios 
$R_{n,L}^{B(E2)}$ and $\underline{R}_{n,L}^{B(E2)}$), while at the left hand 
side the X(5)-$\beta^2$ values, given in Ref. \cite{X5}, are obtained.  

In Table 3 interband B(E2) transition rates from the $n=1$ band to the 
$n=0$ band (defined in Eq. (\ref{eq:e41})) 
are also shown. These transitions are forbidden 
in the SU(3) framework \cite{IA}. The rapid fall of these transitions towards 
zero can be seen both in Table 3 and in Fig. 6(b), where the X(5)-$\beta^2$ 
limiting values on the left are in agreement with the ones given in 
Ref. \cite{X5}. 

Furthermore in Table 3 ratios of quadrupole moments within the $n=0$ and $n=1$ 
bands are given. In the SU(3) limit (for infinite number of bosons) one has 
\cite{IA} for all values of $n$
\begin{equation}\label{eq:e48}
Q_{n,L}= a {L\over 2L+3}, 
\end{equation} 
where $a$ constant, 
corresponding to 
\begin{equation}\label{eq:e49} 
\underline{R}^Q_{n,L} = {7\over 2} {L\over 2L+3}.
\end{equation}
It is clear from Table 3 that at $\beta_0=4$ some differences between the 
$n=0$ and $n=1$ cases are still visible, while at $\beta_0=8$ both cases 
become almost identical to the SU(3) values. 

The evolution of quadrupole moments from the X(5)-$\beta^2$ case 
to the SU(3) limiting values is depicted in Fig. 6(c). 
The SU(3) limiting values
on the right are in good agreement with the contents of Table 3
(taking into account the difference between the ratios $R^Q_{n,L}$ and 
$\underline{R}^Q_{n,L}$, defined in Eqs. (\ref{eq:e43}) and (\ref{eq:e44})). 
Since no results for quadrupole moments for the X(5)-$\beta^2$ case 
are given in Ref. \cite{X5}, they are reported here in Table 4. 
Quadrupole moments for the X(5) model, as well as for the X(5)-$\beta^4$, 
X(5)-$\beta^6$, and X(5)-$\beta^8$ models (defined in Ref. \cite{X5}) 
are also listed in Table 4 as a by-product. 

It should be noted that the SU(3) limit of IBM (at large boson numbers)
is also obtained \cite{FortX5} 
by using in the X(5) framework the exactly soluble Kratzer potential
\cite{Kratzer} 
of Eq. (\ref{eq:e24a}). The SU(3) limit is obtained for large values of $B$. 
In the case of the Kratzer potential, however, it is clear that small values 
of $B$ lead \cite{FortX5} to the Coulomb potential. 

{\bf 3.3 Variational procedure applied to energy ratios } 

The variational procedure used in subsection 2.3 can also be applied here. 
Wishing to determine the critical point in the shape phase transition 
from U(5) to SU(3), one chooses a potential (the Davidson potential
of Eq. (\ref{eq:e1})) 
with a free parameter ($\beta_0$), which serves in spanning the range 
of interest. For large values of $\beta_0$ the SU(3) limit of IBM 
(with large boson numbers) is obtained, 
while for $\beta_0=0$ the X(5)-$\beta^2$ picture occurs \cite{X5}, 
which is not the U(5) limit, but it is located between U(5) and X(5), 
on the way from U(5) to SU(3). Thus the region of interest around X(5) 
is covered from X(5)-$\beta^2$ to SU(3).  

Then the values of $\beta_0$ at which the first derivative $dR_L/d\beta_0$
exhibits a sharp maximum, as seen in the upper panel of Fig. 5,  
are determined for each value of the angular 
momentum $L$ separately. Numerical results for $\beta_{0,m}$ are shown 
in Table 5, together with 
the values of $R_L$ occuring at these points, which are compared to the 
$R_L$ ratios occuring in the ground state band of the X(5) model \cite{IacX5}. 
Very close agreement of the values determined by the variational procedure  
with the X(5) results is observed, thus indicating that the choice of the 
infinite well potential used in the X(5) model is the optimum one for 
the description of the shape phase transition from U(5) to SU(3).  

The results are depicted in Fig. 7(a), where in addition to the bands 
provided by the variational procedure and the X(5) model, the bands 
corresponding to the U(5) and SU(3) cases, calculated from Eqs. 
(\ref{eq:e22}) and (\ref{eq:e45}), as well as the band corresponding to 
X(5)-$\beta^2$ (taken from Ref. \cite{X5}) are shown. 
Finally, the potentials obtained for different angular momenta $L$ 
are shown in Fig. 8, which looks very similar to Fig. 4, obtained 
in the E(5) framework. 

For excited $K=0$ bands the ratios $\underline{R}_{n,L}$ should be considered. 
However, 
because of Eq. (\ref{eq:e30}), it is clear that all these ratios 
for all $K=0$ bands are identical 
to $R_L$ for all values of $n$. This feature is due to the oscillator-like
form of the spectrum of Davidson potentials and is lifted when using the 
generalized Davidson potentials of Eq. (\ref{eq:e25}), which will be further 
discussed in Section 4.  

For $K=2$ bands the ratios $\underline{R'}_{n,L}$ (defined in 
Eq. (\ref{eq:e37a})) should be used. The relevant results 
for the $n=0$, $K=2$ band are reported in 
Table 5 and Fig. 7(b). Not only the energy ratios obtained through the 
variational procedure are very close to the X(5) results \cite{X5,Bijker}, 
but in addition the $\beta_{0,m}$ values obtained for the even values of $L$ 
are very close to the corresponding ones obtained from the application 
of the variational procedure to the ground state band, discussed above. 

{\bf 3.4 Variational procedure applied to excitation energies} 

As in subsection 2.4, it is instructive to apply the variational procedure 
to the excitation energies of the levels of the ground state band, given by 
Eq. (\ref{eq:e31}). Numerical results are shown in Table 5, and are seen 
to be very close to the results provided by the X(5)-$\beta^2$ model \cite{X5}
(which corresponds to a Davidson potential with $\beta_0=0$).   

This result is again an indication that the method works efficiently. 
Indeed, as mentioned in subsection 2.4, 
it is well known \cite{Sakai} that excitation energies 
within a series of isotopes drop very rapidly as one moves away from the 
 vibrational behaviour towards the rotational region, where the change 
is slow. Trying then to identify a series of energy levels corresponding to 
the most rapid changes in the excitation energies, one naturally ends 
up with the most vibrational behaviour possible within the realm 
of the model used, i.e. with the X(5)-$\beta^2$ model in the present case.  

{\bf 3.5 Variational procedure applied to B(E2) ratios} 

As explained in subsection 2.5, the ratios $\underline{R}_{n,L}^{B(E2)}$ 
should be considered in this case. Numerical results for the ground state band
($n=0$, $K=0$) and the $n=1$, $K=0$ band are given in Table 5, 
and depicted in Figs. 7(c) and 7(d), where the U(5) and SU(3) results, 
calculated from Eqs. (\ref{eq:e23}), (\ref{eq:e24}) and (\ref{eq:e47}) 
respectively, are shown
for comparison. In addition, the results given by the original X(5) model 
\cite{IacX5}, as well as by the X(5)-$\beta^2$ and X(5)-$\beta^4$ models
\cite{X5}, which use the $u(\beta)=\beta^2/2$ and  $u(\beta)=\beta^4/2$ 
potentials in the X(5) framework 
instead of an infinite well potential, are exhibited. In both bands 
the variational procedure leads to results which are quite close 
to the X(5)-$\beta^4$ case. Further discussion of these results 
is defered to Section 4. 

It should be mentioned at this point that the first derivative of the 
$B(E2; L+2\to L)$ values with respect to $\beta_0$ again does not exhibit 
a maximum, while the second derivative does not vanish at any value 
other than $\beta_0=0$. Therefore the variational procedure cannot be applied 
to isolated B(E2) values, being applicable to B(E2) ratios only, as in the 
E(5) framework.  

{\bf 3.6 Variational procedure applied to quadrupole moments} 

The variational procedure can also be applied to quadrupole moments. 
Treating each band as a separate entity, one should use the ratios 
$\underline{R}_{n,L}^Q$, defined in Eq. (\ref{eq:e44}), which involve 
quadrupole 
moments of only one band, in contrast to the ratios $R_{n,L}^Q$, defined in 
Eq. (\ref{eq:e43}), which involve quadrupole moments from two different bands,
except in the case of the ground state band. Numerical results for the 
ground state band ($n=0$) and the $n=1$ band are given in Table 5, 
and plotted in Figs. 7(e) and 7(f), where the SU(3)
results (calculated from Eq. (\ref{eq:e49})) are shown for comparison. 
In addition, the results provided by the X(5), X(5)-$\beta^2$ and 
X(5)-$\beta^4$ 
models, given in Table 4, are shown. In both cases it is clear that 
the results of the variational procedure are close to the X(5)-$\beta^4$ 
values. These results will be further discussed in the next section. 

{\bf 4. Discussion}

The main results obtained in the present work are summarized here: 

1) A variational procedure for determining the values of physical 
quantities at the point of shape phase transitions in nuclei 
has been suggested. Using one-parameter potentials spanning the region 
between the two limiting symmetries of interest, the parameter values 
at which the rate of change of the physical quantity becomes maximum are 
determined for each value of the angular momentum separately and the 
corresponding values of the physical quantity at these parameter 
values are calculated. The values of the physical quantity collected in this 
way represent its behavior at the critical point.

2) The method has been applied in the shape phase transition from U(5) to O(6),
using one-parameter Davidson potentials \cite{Dav} and considering 
the energy ratios $R_L=E(L)/E(2)$ within the ground state band as the 
relevant physical quantity, leading to a band which practically coincides 
with the ground state band of the E(5) model \cite{IacE5}. 
It has also been applied in the same way in the shape phase transition from 
U(5) to SU(3), leading to a band which practically coincides with 
the ground state band of the X(5) model \cite{IacX5}. Energy ratios within
the lowest $K=2$ band in the latter case also lead to the relevant X(5) results
\cite{X5,Bijker}. 

3) The method has also been applied to intraband B(E2) ratios of the ground 
state band and the first excited band in the U(5)-O(6) transition region,
leading to results very close to the ones provided by the E(5)-$\beta^4$ 
model \cite{Arias,E5}, which uses a $u(\beta)=\beta^4/2$ potential instead 
of an infinite well potential in the E(5) framework.  
It has also been applied to intraband 
B(E2) ratios and ratios of quadrupole moments of the ground state band 
and the first excited band in the U(5)-SU(3) transition region, leading 
to results very similar to the ones provided by the X(5)-$\beta^4$ model
\cite{X5}, which uses a $u(\beta)=\beta^4/2$ potential instead of an 
infinite well potential in the X(5) framework. 

4) The method has also been applied to isolated excitation energies 
of the ground state band, leading to a U(5) band in the E(5) framework and to 
a X(5)-$\beta^2$ \cite{X5} band in the X(5) framework. 
(The U(5) and X(5)-$\beta^2$ models correspond to the use of a harmonic 
oscillator potential $u(\beta)=\beta^2/2$ in the E(5) and X(5) frameworks,
respectively.) 
These results are 
expected, since it is known \cite{Sakai} that the most rapid change (drop) 
of the excitation energies in a series of isotopes occurs as one 
starts moving away from the vibrational limit towards the rotational limit. 

5) It should be emphasized that the application of the method was possible 
because the Davidson potentials correctly reproduce the U(5) and O(6) 
symmetries of IBM (with large boson numbers) in the E(5) framework (for small 
and large parameter values 
respectively), as well as the relevant X(5)-$\beta^2$ \cite{X5} and SU(3) 
symmetries in the X(5) framework (for small and large parameter values 
respectively). The occurence of SU(3) (with large boson numbers)
in the X(5) framework is a new result, 
which has been proved by considering energy, intraband and interband B(E2),
and quadrupole moment ratios, while the occurence of O(6) in the E(5) 
framework has essentially been observed earlier \cite{Elliott} and 
has been corroborated here by considering energy and intraband B(E2) ratios.  

6) As a by-product, a derivation of the Holmberg--Lipas formula \cite{Lipas} 
has been achieved using Davidson potentials in the X(5) framework. 

7) As another by-product, quadrupole moments for the X(5) model and the 
X(5)-$\beta^{2n}$ models \cite{X5} for $n=1$, 2, 3, 4 have been calculated. 

The following comments are now in place: 

i) The fact that the application of the variational procedure to energy ratios 
within the ground state band in the E(5) and X(5) frameworks leads to results 
very close to the ground state bands of the E(5) and X(5) models suggests that
the selection of the infinite well potential is the optimum one in both cases. 

ii) The fact that the application of the variational procedure to intraband 
B(E2) transition ratios and to quadrupole moment ratios leads to results 
close to the E(5)-$\beta^4$ \cite{Arias,E5} and X(5)-$\beta^4$ \cite{X5} 
models in the E(5) and X(5) frameworks respectively, suggests that further 
studies are needed. In particular, it is of interest to apply the variational 
procedure using the generalized Davidson potentials of Eq. (\ref{eq:e25}) 
as ``trial potentials'' in both the E(5) and X(5) frameworks. These potentials 
with $\beta_0=0$ are known to approach smoothly the E(5) and X(5) models 
from the U(5) direction \cite{E5,X5}, as the power of $\beta$ in the 
$\beta^{2n}$ term increases. It is expected that these potentials 
with $\beta_0\neq 0$ will be smoothly approaching the E(5) and X(5) models 
from the O(6) and SU(3) directions, respectively. Then the results 
of the variational procedure could converge towards the E(5) and X(5) 
results with increasing $n$. 

iii) The application of the variational procedure to energy ratios involving 
levels of excited bands with $n>0$ is trivial in the case of the Davidson 
potential, because of its harmonic oscillator features, but it becomes 
nontrivial in the case of the generalized Davidson potentials, and therefore 
this task should be undertaken.   

iv) The generalized Davidson potentials are known to possess the appropriate 
limiting behaviour for $\beta_0=0$ \cite{E5,X5} and are expected to approach 
the appropriate limits (near to O(6) and SU(3) in the E(5) and X(5) frameworks
respectively) for large values of $\beta_0$. Any other 
potential/Hamiltonian bridging the relevant pairs of symmetries 
(U(5)-O(6) and U(5)-SU(3)) could be equally appropriate. 

v) It is interesting that the most general (up to two-body terms) IBM 
Hamiltonian appropriate for the U(5) to O(6) transition leads \cite{Arias}
to the E(5)-$\beta^4$ model, in agreement to the results mentioned in ii). 
It will also be interesting to examine if appropriate symmetry-conserving 
higher order terms \cite{Chen,Heyde,Berghe,Vanth}, 
when added to this Hamiltonian, modify this conclusion. 

{\bf Acknowledgements}

Partial support through the NATO Collaborative Linkage Grant PST.CLG 978799 is 
gratefully acknowledged.

\newpage 

\centerline{\bf Figure captions} 

{\bf Fig. 1} (Color online)
$R_L$ energy ratios (Eq. (\ref{eq:e14})) for the ground state 
band (for $L=4$, 12, 20) and their derivatives $dR_L/d\beta_0$ vs. the 
parameter $\beta_0$, calculated using Davidson potentials (Eq. (\ref{eq:e1})) 
in the E(5) framework. The $R_L$ curves demonstrate the evolution from the 
U(5) symmetry (on the left) to the O(6) limit of IBM with large boson numbers 
(on the right). See subsections 2.2 and 2.3 for further details. 

{\bf Fig. 2} (Color online)
Intraband B(E2) ratios (Eq. (\ref{eq:e17})) $R_{0,L}^{B(E2)}$ 
(for the ground state band) and $R_{1,L}^{B(E2)}$ (for the $n=1$ band)  
for $L=2$, 10, 18, vs. the parameter $\beta_0$. The curves show the 
evolution from the U(5) symmetry (on the left) to the O(6) limit 
of IBM with large boson numbers (on the right). See subsections 2.2 and 
2.5 for further discussion. 

{\bf Fig. 3} (Color online)
Energy ratios $R_L$ [Eq. (\ref{eq:e14})] (a) and 
intraband B(E2) ratios $\underline{R}^{B(E2)}_{0,L}$ [Eq. (\ref{eq:e18})] (b) 
for the ground state band, as well as intraband B(E2) ratios 
$\underline{R}^{B(E2)}_{1,L}$ [Eq. (\ref{eq:e18})] (c) for the $n=1$ band, 
obtained through the variational procedure (labeled by ``var'') using 
Davidson potentials in the E(5) framework, compared to the values provided 
by the U(5), O(6), E(5), and E(5)-$\beta^4$ models. 
See subsections 2.3 and 2.5 for further details. 

{\bf Fig. 4} (Color online) 
Davidson potentials (Eq. (\ref{eq:e1})) obtained for different 
angular momenta $L$ through the application of the variational procedure to 
energy ratios within the ground state band in the E(5) framework.
The $\beta_0$ values corresponding to these potentials are listed 
in Table 2. See subsection 2.3 for further discussion.  

{\bf Fig. 5} (Color online) 
$R_L$ energy ratios (Eq. (\ref{eq:e14})) for the ground state 
band (for $L=4$, 12, 20) and their derivatives $dR_L/d\beta_0$ vs. the 
parameter $\beta_0$, calculated using Davidson potentials (Eq. (\ref{eq:e1})) 
in the X(5) framework. The $R_L$ curves demonstrate the evolution from the 
X(5)-$\beta^2$ symmetry (on the left) to the SU(3) limit of IBM with large 
boson numbers (on the right). See subsections 3.2 and 3.3 for further details. 

{\bf Fig. 6} (Color online) 
(a) Intraband B(E2) ratios (Eq. (\ref{eq:e17})) $R_{0,L}^{B(E2)}$
(for the ground state band) and $R_{1,L}^{B(E2)}$ (for the $n=1$, $K=0$ band) 
for $L=2$, 10, 18,  vs. the parameter $\beta_0$. (b) Interband B(E2) ratios 
$R_{1,L,0,L'}^{B(E2)}$ (Eq. (\ref{eq:e41})) vs. $\beta_0$. (c) Quadrupole 
moment ratios (Eq. (\ref{eq:e43}))$R_{0,L}^Q$ (for the ground state band) and 
$R_{1,L}^Q$ (for the $n=1$, $K=0$ band) for $L=4$, 12, 20, vs. $\beta_0$. 
In all cases the curves show the evolution from the X(5)-$\beta^2$ model (on 
the left) to the SU(3) limit of IBM with large boson numbers (on the right). 
See subsections 3.2, 3.5, and 3.6 for further discussion. 

{\bf Fig. 7} (Color online) 
Energy ratios $R_L$ [Eq. (\ref{eq:e14})] (a), 
intraband B(E2) ratios $\underline{R}^{B(E2)}_{0,L}$ [Eq. (\ref{eq:e18})] (c),
and quadrupole moment ratios $\underline{R}^Q_{0,L}$ [Eq. (\ref{eq:e44})] (e) 
for the ground state band, as well as intraband B(E2) ratios 
$\underline{R}^{B(E2)}_{1,L}$ [Eq. (\ref{eq:e18})] (d) and quadrupole moment 
ratios $\underline{R}^Q_{1,L}$ [Eq. (\ref{eq:e44})] (f) for the $n=1$, $K=0$ 
band, and energy ratios $\underline{R'}_{0,L}$ [Eq. (\ref{eq:e37a})] (b)
for the $n=0$, $K=2$ band
obtained through the variational procedure (labeled by ``var'') using 
Davidson potentials in the X(5) framework, compared to the values provided 
by the U(5), SU(3), X(5), X(5)-$\beta^2$ and X(5)-$\beta^4$ models. 
See subsections 3.3, 3.5 and 3.6 for further details. 

{\bf Fig. 8} (Color online)
Davidson potentials (Eq. (\ref{eq:e1})) obtained for different 
angular momenta $L$ through the application of the variational procedure to 
energy ratios within the ground state band in the X(5) framework.
The $\beta_0$ values corresponding to these potentials are listed 
in Table 5. See subsection 3.3 for further discussion.  

\newpage 
\parindent=0pt

\begin{table}

\caption{$R_L$ energy ratios (Eq. (\ref{eq:e14})) and 
$\underline{R}_{0,L-2}^{B(E2)}$ intraband B(E2) ratios (Eq. (\ref{eq:e18})) 
for the ground state band, as well as $\underline{R}_{1,L-2}^{B(E2)}$ intraband
B(E2) ratios (Eq. (\ref{eq:e18})) for the $n=1$ band of the Davidson potentials
in the E(5) framework for different values of the parameter $\beta_0$, 
compared to the O(6) results for large boson numbers. 
See subsection 2.2 for further details.  
}

\bigskip

\begin{tabular}{r r r r | r r r | r r r}
\hline
$L$  &  $R_L$ & $R_L$ & $R_L$ & $\underline{R}_{0,L-2}^{B(E2)}$ & 
$\underline{R}_{0,L-2}^{B(E2)}$ & $\underline{R}_{0,L-2}^{B(E2)}$  & 
$\underline{R}_{1,L-2}^{B(E2)}$ & $\underline{R}_{1,L-2}^{B(E2)}$ & 
$\underline{R}_{1,L-2}^{B(E2)}$ \\
   & $\beta_0=4.$ & $\beta_0= 8.$ & O(6) & $\beta_0=4.$ & $\beta_0=8.$ & O(6) 
& $\beta_0=4.$ & $\beta_0=8.$ & O(6) \\
\hline
2  &  1.000 &  1.000 &  1.000 & 1.000 & 1.000 & 1.000 & 1.000 & 1.000 & 1.000\\
4  &  2.486 &  2.499 &  2.500 & 1.441 & 1.429 & 1.429 & 1.439 & 1.429 & 1.429\\
6  &  4.441 &  4.496 &  4.500 & 1.702 & 1.669 & 1.667 & 1.697 & 1.669 & 1.667\\
8  &  6.846 &  6.990 &  7.000 & 1.885 & 1.823 & 1.818 & 1.876 & 1.823 & 1.818\\
10 &  9.677 &  9.978 & 10.000 & 2.030 & 1.930 & 1.923 & 2.015 & 1.930 & 1.923\\
12 & 12.909 & 13.459 & 13.500 & 2.154 & 2.011 & 2.000 & 2.133 & 2.010 & 2.000\\
14 & 16.515 & 17.430 & 17.500 & 2.268 & 2.074 & 2.059 & 2.240 & 2.073 & 2.059\\
16 & 20.469 & 21.888 & 22.000 & 2.376 & 2.125 & 2.105 & 2.341 & 2.124 & 2.105\\
18 & 24.743 & 26.831 & 27.000 & 2.482 & 2.167 & 2.143 & 2.438 & 2.166 & 2.143\\
20 & 29.311 & 32.254 & 32.500 & 2.587 & 2.204 & 2.174 & 2.535 & 2.203 & 2.174\\
\hline 
\end{tabular}
\end{table}

\vskip 5.0truein

\newpage 
\parindent=0pt

\begin{table}

\caption{Parameter values $\beta_{0,m}$ where the absolute value of the first 
derivative of a physical quantity has a maximum, while the second derivative 
vanishes, for the $R_L$ energy ratios (Eq. (\ref{eq:e14})), the 
$\underline{R}_{0,L-2}^{B(E2)}$ intraband B(E2) ratios (Eq. (\ref{eq:e18})), 
and the $E_{0,L,exc}$ excitation energies (Eq. (\ref{eq:e9})) of the 
ground state band, as well as for the $\underline{R}_{1,L-2}^{B(E2)}$ intraband
B(E2) ratios (Eq. (\ref{eq:e18})) for the $n=1$ band of the Davidson potentials
in the E(5) framework, together with the corresponding values of 
each physical quantity (labeled by ``var''), which are compared to 
appropriate E(5), E(5)-$\beta^4$, or U(5) results. In the case of 
$E_{0,L,exc}$, the corresponding $R_L$ ratios (Eq. (\ref{eq:e14})) 
are also shown. See subsections 2.3, 2.4, and 2.5 for further details. 
}

\bigskip

\begin{tabular}{r r r r | r r r | r r r}
\hline
$L$ & $\beta_{0,m}$ & $R_L$ & $R_L$ & 
$\beta_{0,m}$ & $\underline{R}_{0,L-2}^{B(E2)}$ & 
$\underline{R}_{0,L-2}^{B(E2)}$ & 
$\beta_{0,m}$ & $\underline{R}_{1,L-2}^{B(E2)}$ & 
$\underline{R}_{1,L-2}^{B(E2)}$ \\
  &  & var & E(5) & &  var  &  E(5)-$\beta^4$ & & var & E(5)-$\beta^4$ \\
\hline
4  & 1.421 &  2.185 &  2.199 & 1.213 & 1.801 & 1.832 & 1.124 & 1.697 & 1.757 \\
6  & 1.522 &  3.549 &  3.590 & 1.266 & 2.538 & 2.564 & 1.184 & 2.300 & 2.406 \\
8  & 1.609 &  5.086 &  5.169 & 1.299 & 3.252 & 3.227 & 1.224 & 2.868 & 2.990 \\
10 & 1.687 &  6.793 &  6.934 & 1.321 & 3.960 & 3.841 & 1.250 & 3.421 & 3.529 \\
12 & 1.759 &  8.667 &  8.881 & 1.336 & 4.667 & 4.417 & 1.269 & 3.967 & 4.035 \\
14 & 1.825 & 10.705 & 11.009 & 1.347 & 5.375 & 4.961 & 1.283 & 4.510 & 4.513 \\
16 & 1.888 & 12.906 & 13.316 & 1.355 & 6.085 & 5.480 & 1.293 & 5.051 & 4.969 \\
18 & 1.947 & 15.269 & 15.799 & 1.361 & 6.798 & 5.978 & 1.300 & 5.593 & 5.407 \\
20 & 2.004 & 17.793 & 18.459 & 1.365 & 7.511 & 6.457 & 1.306 & 6.135 & 5.828 \\
\hline 
\end{tabular}
\end{table}

\newpage 
\parindent=0pt
\setcounter{table}{1} 
\begin{table}

\caption{(continued)}

\bigskip

\begin{tabular}{r r r r r}
\hline
$L$ & $\beta_{0,m}$ & $E_{0,L,exc}$ &  $R_L$ & $R_L$ \\
  &   & var &  var &  U(5)  \\
\hline
2  & 1.399 & 0.709 &  1.000 &  1.000 \\
4  & 1.538 & 1.423 &  2.008 &  2.000 \\
6  & 1.658 & 2.142 &  3.021 &  3.000 \\
8  & 1.766 & 2.862 &  4.037 &  4.000 \\
10 & 1.866 & 3.583 &  5.054 &  5.000 \\
12 & 1.960 & 4.303 &  6.070 &  6.000 \\
14 & 2.049 & 5.022 &  7.084 &  7.000 \\
16 & 2.135 & 5.739 &  8.096 &  8.000 \\
18 & 2.217 & 6.455 &  9.105 &  9.000 \\
20 & 2.297 & 7.168 & 10.111 & 10.000 \\
\hline 
\end{tabular}
\end{table}

\newpage 
\parindent=0pt

\begin{table}

\caption{$R_L$ energy ratios (Eq. (\ref{eq:e14})), 
 $\underline{R}_{0,L-2}^{B(E2)}$ intraband B(E2) ratios (Eq. (\ref{eq:e18})),
and $\underline{R}_{0,L}^Q$ quadrupole moment ratios (Eq. (\ref{eq:e44})) 
for the ground state band, as well as $\underline{R}_{1,L-2}^{B(E2)}$ 
intraband B(E2) ratios (Eq. (\ref{eq:e18})) and $\underline{R}_{1,L}^Q$ 
quadrupole moment ratios (Eq. (\ref{eq:e44})) for the $n=1$, $K=0$  band, 
and $\underline{R'}_{0,L}$ energy ratios (Eq. (\ref{eq:e37a}))
for the $n=0$, $K=2$ band of the Davidson potentials 
in the X(5) framework for different values of the parameter 
$\beta_0$, compared to the SU(3) results at large boson numbers. 
In addition, $R^{B(E2)}_{1,L,0,L'}$ interband B(E2) ratios (Eq. (\ref{eq:e41})
between the $n=1$, $K=0$ band and the ground state band are given.  
See subsection 3.2 for further details.  }

\bigskip

\begin{tabular}{r r r r | r r r | r r r}
\hline
$L$  &  $R_L$ & $R_L$ & $R_L$ & 
$\underline{R}_{0,L-2}^{B(E2)}$ & $\underline{R}_{0,L-2}^{B(E2)}$ & 
$\underline{R}_{0,L-2}^{B(E2)}$ & $\underline{R}_{1,L-2}^{B(E2)}$ & 
$\underline{R}_{1,L-2}^{B(E2)}$ & $\underline{R}_{1,L-2}^{B(E2)}$
\\
   & $\beta_0=4.$ & $\beta_0=8.$ & SU(3) & 
     $\beta_0=4.$ & $\beta_0=8.$ & SU(3) &
     $\beta_0=4.$ & $\beta_0=8.$ & SU(3) \\
\hline
2  &  1.000 &  1.000 &  1.000 & 1.000 & 1.000 & 1.000 & 1.000 & 1.000 & 1.000\\
4  &  3.318 &  3.332 &  3.333 & 1.437 & 1.429 & 1.429 & 1.436 & 1.429 & 1.429\\
6  &  6.921 &  6.995 &  7.000 & 1.599 & 1.575 & 1.573 & 1.595 & 1.575 & 1.573\\
8  & 11.756 & 11.984 & 12.000 & 1.699 & 1.651 & 1.647 & 1.690 & 1.650 & 1.647\\
10 & 17.759 & 18.295 & 18.333 & 1.777 & 1.698 & 1.692 & 1.764 & 1.697 & 1.692\\
12 & 24.856 & 25.921 & 26.000 & 1.849 & 1.731 & 1.722 & 1.829 & 1.730 & 1.722\\
14 & 32.967 & 34.856 & 35.000 & 1.919 & 1.756 & 1.743 & 1.892 & 1.755 & 1.743\\
16 & 42.011 & 45.091 & 45.333 & 1.991 & 1.776 & 1.760 & 1.957 & 1.775 & 1.760\\
18 & 51.904 & 56.616 & 57.000 & 2.066 & 1.793 & 1.772 & 2.023 & 1.793 & 1.772\\
20 & 62.570 & 69.421 & 70.000 & 2.143 & 1.809 & 1.782 & 2.092 & 1.808 & 1.782\\
\hline 
\end{tabular}
\end{table}

\newpage 
\parindent=0pt
\setcounter{table}{2}

\begin{table}

\caption{(continued) 
}

\bigskip

\begin{tabular}{r r r r | r r r r | r r r}
\hline
$L$ & $L'$ & $100 R_{1,L,0,L'}^{B(E2)}$ & $100 R_{1,L,0,L'}^{B(E2)}$ & $L$ & 
$\underline{R}^Q_{0,L}$ & $\underline{R}^Q_{0,L}$ & $\underline{R}^Q_{0,L}$ & 
$\underline{R}^Q_{1,L}$ & $\underline{R}^Q_{1,L}$ & $\underline{R}^Q_{1,L}$ \\
 &  & $\beta_0=4.$ & $\beta_0= 8.$ &  & 
$\beta_0=4.$ & $\beta_0=8.$ & SU(3) & 
$\beta_0=4.$ & $\beta_0=8.$ & SU(3) \\
\hline
 0 & 2 &  8.261 &  1.983 & 2 & 1.000 & 1.000 & 1.000 & 1.000 & 1.000& 1.000 \\
 2 & 0 &  1.284 &  0.373 & 4 & 1.278 & 1.273 & 1.273 & 1.278 & 1.273& 1.273 \\
 2 & 2 &  2.089 &  0.549 & 6 & 1.415 & 1.401 & 1.400 & 1.414 & 1.401& 1.400 \\
 2 & 4 &  4.940 &  1.062 & 8 & 1.503 & 1.476 & 1.474 & 1.500 & 1.476& 1.474 \\
 4 & 2 &  1.524 &  0.510 &10 & 1.568 & 1.525 & 1.522 & 1.564 & 1.525& 1.522 \\
 4 & 4 &  1.899 &  0.499 &12 & 1.622 & 1.560 & 1.556 & 1.617 & 1.560& 1.556 \\
 4 & 6 &  5.006 &  0.977 &14 & 1.671 & 1.587 & 1.581 & 1.664 & 1.587& 1.581 \\
 6 & 4 &  1.377 &  0.537 &16 & 1.716 & 1.608 & 1.600 & 1.707 & 1.608& 1.600 \\ 
 6 & 6 &  1.861 &  0.489 &18 & 1.760 & 1.626 & 1.615 & 1.749 & 1.626& 1.615 \\
 6 & 8 &  5.365 &  0.963 &20 & 1.803 & 1.641 & 1.628 & 1.789 & 1.641& 1.628 \\
\hline 
\end{tabular}
\end{table}

\newpage 
\parindent=0pt
\setcounter{table}{2}

\begin{table}

\caption{(continued) 
}

\bigskip

\begin{tabular}{r r r r }
\hline
$L$  &  $\underline{R'}_{0,L}$ & $\underline{R'}_{0,L}$ & 
$\underline{R'}_{0,L}$ \\ 
   & $\beta_0=4.$ & $\beta_0=8.$ & SU(3) \\ 
\hline
3  &  1.000 &  1.000 &  1.000 \\
4  &  2.327 &  2.333 &  2.333 \\
5  &  3.977 &  3.999 &  4.000 \\
6  &  5.944 &  5.996 &  6.000 \\
7  &  8.219 &  8.326 &  8.333 \\
8  & 10.797 & 10.987 & 11.000 \\
9  & 13.667 & 13.978 & 14.000 \\
10 & 16.821 & 17.299 & 17.333 \\
\hline 
\end{tabular}
\end{table}

\newpage 
\parindent=0pt

\begin{table}

\caption{Quadrupole moments (Eq. (\ref{eq:e42})) of the X(5)-$\beta^4$, 
X(5)-$\beta^6$, and X(5)-$\beta^8$ models, compared to the predictions 
of the X(5) and X(5)-$\beta^2$ models for some $K=0$ bands. 
See subsection 3.1 for details. 
}

\bigskip

\begin{tabular}{r r r r r r r} 
\hline
band & $L$ & X(5)-$\beta^2$ & X(5)-$\beta^4$ & X(5)-$\beta^6$ & 
X(5)-$\beta^8$ & X(5) \\
\hline
$s=1$ & & & & &   &       \\
    & 0  & 0.000 & 0.000 & 0.000 & 0.000 & 0.000 \\
    & 2  & 1.000 & 1.000 & 1.000 & 1.000 & 1.000 \\
    & 4  & 1.466 & 1.413 & 1.391 & 1.380 & 1.358 \\
    & 6  & 1.823 & 1.699 & 1.648 & 1.622 & 1.572 \\
    & 8  & 2.127 & 1.925 & 1.842 & 1.800 & 1.719 \\
    & 10 & 2.395 & 2.114 & 2.000 & 1.941 & 1.828 \\
    & 12 & 2.638 & 2.277 & 2.132 & 2.058 & 1.913 \\
    & 14 & 2.862 & 2.422 & 2.247 & 2.158 & 1.982 \\
    & 16 & 3.070 & 2.552 & 2.348 & 2.245 & 2.038 \\
    & 18 & 3.266 & 2.671 & 2.439 & 2.322 & 2.086 \\
    & 20 & 3.450 & 2.781 & 2.522 & 2.391 & 2.127 \\
    & 22 & 3.626 & 2.884 & 2.598 & 2.454 & 2.162 \\
    & 24 & 3.794 & 2.980 & 2.669 & 2.512 & 2.193 \\
    & 26 & 3.954 & 3.070 & 2.734 & 2.565 & 2.221 \\
    & 28 & 4.109 & 3.155 & 2.795 & 2.615 & 2.245 \\ 
    & 30 & 4.258 & 3.236 & 2.853 & 2.661 & 2.268 \\ 
\hline
$s=2$ & & & & & &        \\
    & 0 & 0.000 & 0.000 & 0.000 & 0.000 & 0.000\\
    & 2 & 1.245 & 1.107 & 1.034 & 0.991 & 0.894\\
    & 4 & 1.742 & 1.521 & 1.411 & 1.346 & 1.197\\
    & 6 & 2.095 & 1.796 & 1.655 & 1.573 & 1.379\\
    & 8 & 2.387 & 2.012 & 1.840 & 1.743 & 1.508\\
    &10 & 2.643 & 2.191 & 1.992 & 1.880 & 1.608\\
    &12 & 2.875 & 2.347 & 2.120 & 1.995 & 1.689\\
    &14 & 3.088 & 2.486 & 2.233 & 2.095 & 1.756\\
    &16 & 3.287 & 2.612 & 2.333 & 2.182 & 1.813\\
    &18 & 3.473 & 2.727 & 2.423 & 2.260 & 1.863\\
    &20 & 3.651 & 2.833 & 2.505 & 2.331 & 1.906\\
\hline 
\end{tabular}
\end{table}

\newpage 
\parindent=0pt

\begin{table}

\caption{Parameter values $\beta_{0,m}$ where the absolute value of the first 
derivative of a physical quantity has a maximum, while the second derivative 
vanishes, for the $R_L$ energy ratios (Eq. (\ref{eq:e14})), the 
$\underline{R}_{0,L-2}^{B(E2)}$ intraband B(E2) ratios (Eq. (\ref{eq:e18})), 
the $E_{0,L,exc}^{(0)}$ excitation energies (Eq. (\ref{eq:e31})), and the 
$\underline{R}_{0,L}^Q$ quadrupole moment ratios (Eq. (\ref{eq:e44})) of the 
ground state band, as well as for the $\underline{R}_{1,L-2}^{B(E2)}$ intraband
B(E2) ratios (Eq. (\ref{eq:e18})) and the $\underline{R}^Q_{1,L}$ quadrupole 
moment ratios (Eq. (\ref{eq:e44})) of the $n=1$, $K=0$ band, 
and for the $\underline{R'}_{0,L}$ energy ratios (Eq. (\ref{eq:e37a}))
of the $n=0$, $K=2$ band of the Davidson potentials 
in the X(5) framework, together with the corresponding values of 
each physical quantity (labeled by ``var''), which are compared to 
appropriate X(5), X(5)-$\beta^2$, or X(5)-$\beta^4$ results. In the case of 
$E_{0,L,exc}$, the corresponding $R_L$ ratios (Eq. (\ref{eq:e14})) 
are also shown. See subsections 3.3, 3.4, 3.5, and 3.6 for further details. 
}

\bigskip

\begin{tabular}{r r r r | r r r | r r r}
\hline
$L$ & $\beta_{0,m}$ & $R_L$ & $R_L$ & $\beta_{0,m}$ & 
$\underline{R}_{0,L-2}^{B(E2)}$ & 
$\underline{R}_{0,L-2}^{B(E2)}$ & $\beta_{0,m}$ & 
$\underline{R}_{1,L-2}^{B(E2)}$ & $\underline{R}_{1,L-2}^{B(E2)}$  \\
  & & var & X(5) &  & var  & X(5)-$\beta^4$ & & var & X(5)-$\beta^4$ \\
 \hline
4  & 1.334 &  2.901 &  2.904 & 1.234 & 1.658 & 1.690 & 1.059 & 1.517 & 1.539 \\
6  & 1.445 &  5.419 &  5.430 & 1.275 & 2.210 & 2.262 & 1.206 & 1.892 & 1.966 \\
8  & 1.543 &  8.454 &  8.483 & 1.312 & 2.771 & 2.799 & 1.289 & 2.264 & 2.382 \\
10 & 1.631 & 11.964 & 12.027 & 1.337 & 3.346 & 3.305 & 1.345 & 2.641 & 2.781 \\
12 & 1.711 & 15.926 & 16.041 & 1.355 & 3.933 & 3.783 & 1.383 & 3.024 & 3.162 \\
14 & 1.785 & 20.330 & 20.514 & 1.367 & 4.530 & 4.237 & 1.411 & 3.413 & 3.527 \\
16 & 1.855 & 25.170 & 25.437 & 1.375 & 5.135 & 4.671 & 1.431 & 3.807 & 3.877 \\
18 & 1.922 & 30.442 & 30.804 & 1.381 & 5.745 & 5.087 & 1.446 & 4.205 & 4.214 \\
20 & 1.985 & 36.146 & 36.611 & 1.386 & 6.361 & 5.489 & 1.458 & 4.606 & 4.539 \\
\hline 
\end{tabular}
\end{table}

\newpage 
\parindent=0pt
\setcounter{table}{4} 

\begin{table}

\caption{(continued)} 

\bigskip

\begin{tabular}{r r r r r | r r r | r r r}
\hline
$L$ & $\beta_{0,m}$ & $E_{0,L,exc}^{(0)}$ &  $R_L$ & $R_L$ & $\beta_{0,m}$ &
$\underline{R}_{0,L}^Q$ & $\underline{R}_{0,L}^Q$ & $\beta_{0,m}$ & 
$\underline{R}_{1,L}^Q$ & $\underline{R}_{1,L}^Q$ \\
  &   & var & var & X(5)-$\beta^2$ & & var & X(5)-$\beta^4$ & & var & 
X(5)-$\beta^4$ \\
\hline
2  & 1.329 & 0.397 &  1.000 &  1.000 &       &       &       &       & \\
4  & 1.470 & 1.055 &  2.655 &  2.646 & 1.401 & 1.395 & 1.413 & 1.467 & 1.354 &
 1.374 \\
6  & 1.604 & 1.804 &  4.540 &  4.507 & 1.472 & 1.670 & 1.699 & 1.551 & 1.582 &
 1.622 \\
8  & 1.726 & 2.591 &  6.517 &  6.453 & 1.524 & 1.894 & 1.925 & 1.616 & 1.761 &
 1.818 \\
10 & 1.840 & 3.394 &  8.539 &  8.438 & 1.562 & 2.090 & 2.114 & 1.668 & 1.914 &
 1.979 \\
12 & 1.947 & 4.206 & 10.582 & 10.445 & 1.591 & 2.267 & 2.277 & 1.708 & 2.051 &
 2.120 \\
14 & 2.049 & 5.022 & 12.634 & 12.465 & 1.613 & 2.431 & 2.422 & 1.741 & 2.176 &
 2.246 \\
16 & 2.146 & 5.839 & 14.690 & 14.494 & 1.630 & 2.584 & 2.552 & 1.767 & 2.294 &
 2.360 \\
18 & 2.240 & 6.656 & 16.745 & 16.529 & 1.644 & 2.730 & 2.671 & 1.789 & 2.405 &
 2.463 \\
20 & 2.330 & 7.472 & 18.798 & 18.568 & 1.654 & 2.868 & 2.781 & 1.806 & 2.511 &
 2.559 \\
\hline 

\end{tabular}
\end{table}

\newpage 
\parindent=0pt
\setcounter{table}{4} 

\begin{table}

\caption{(continued)} 

\bigskip

\begin{tabular}{r r r r}
\hline
$L$ & $\beta_{0,m}$ & $\underline{R'}_{0,L}$ & $\underline{R'}_{0,L}$  \\ 
  & & var & X(5) \\
 \hline
4  & 1.339 &  2.157 &  2.163 \\
5  & 1.404 &  3.454 &  3.472 \\
6  & 1.463 &  4.881 &  4.919 \\
7  & 1.517 &  6.431 &  6.497 \\
8  & 1.567 &  8.101 &  8.205 \\
9  & 1.614 &  9.886 & 10.037 \\
10 & 1.658 & 11.786 & 11.994 \\
\hline 
\end{tabular}
\end{table}

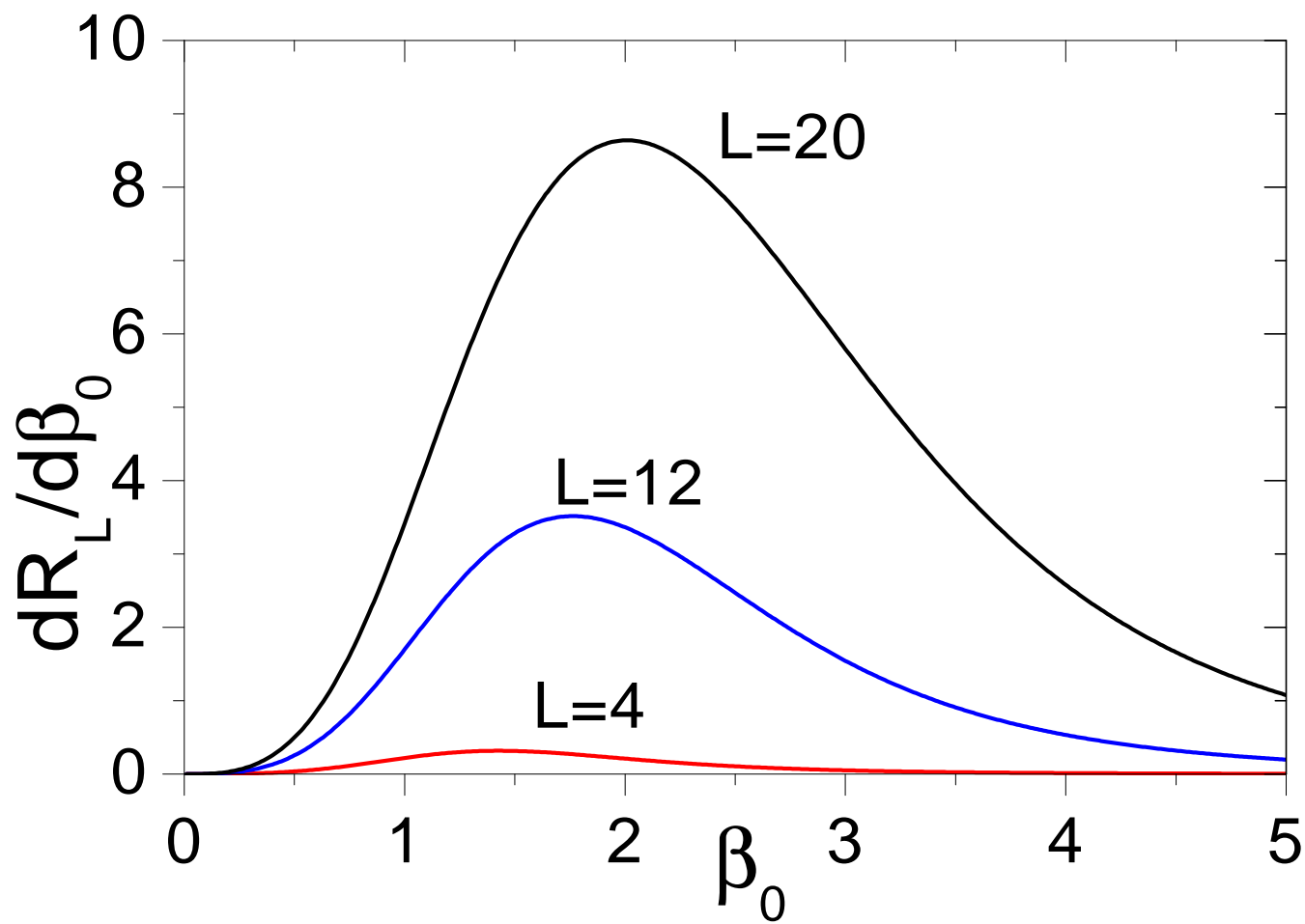
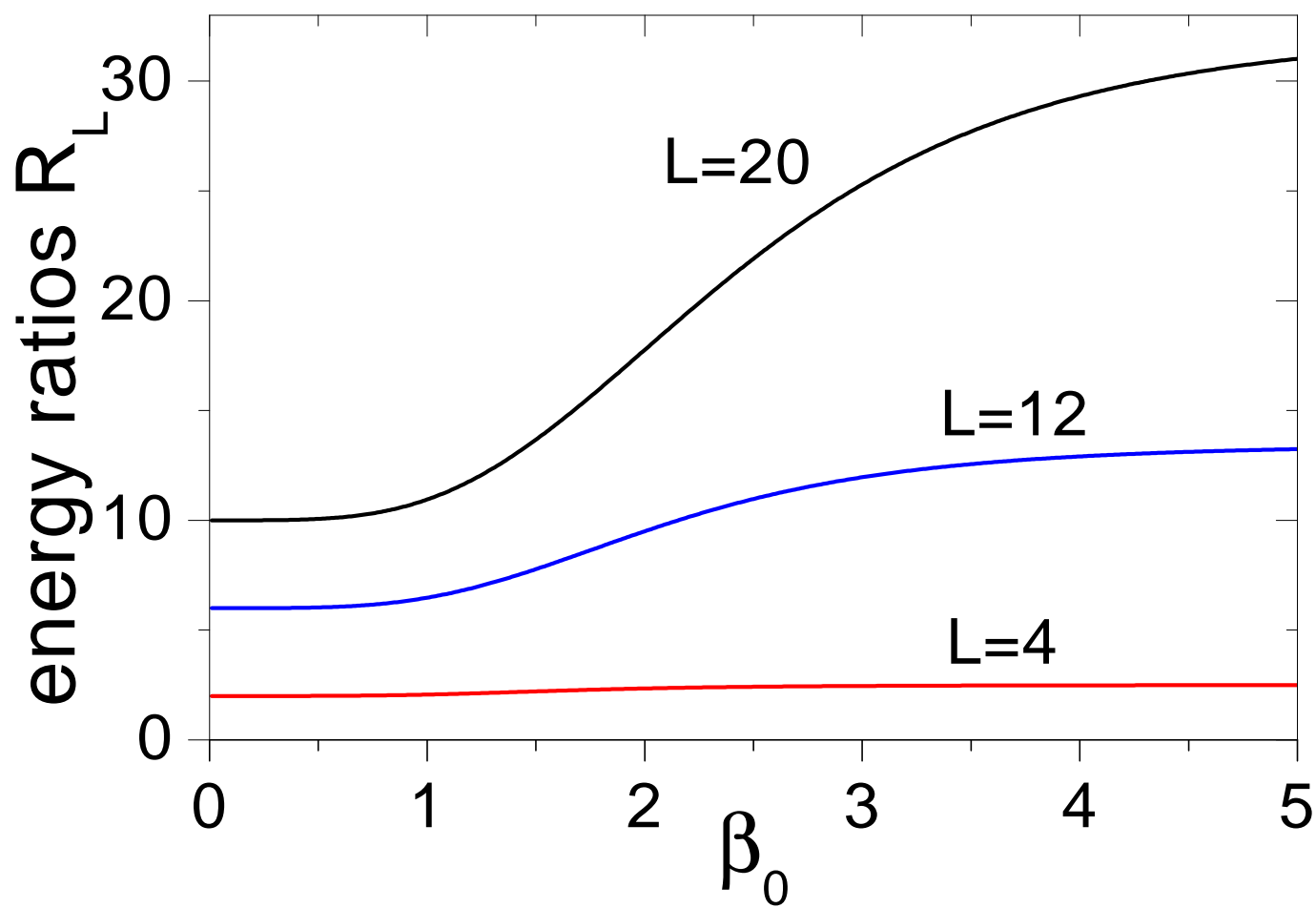
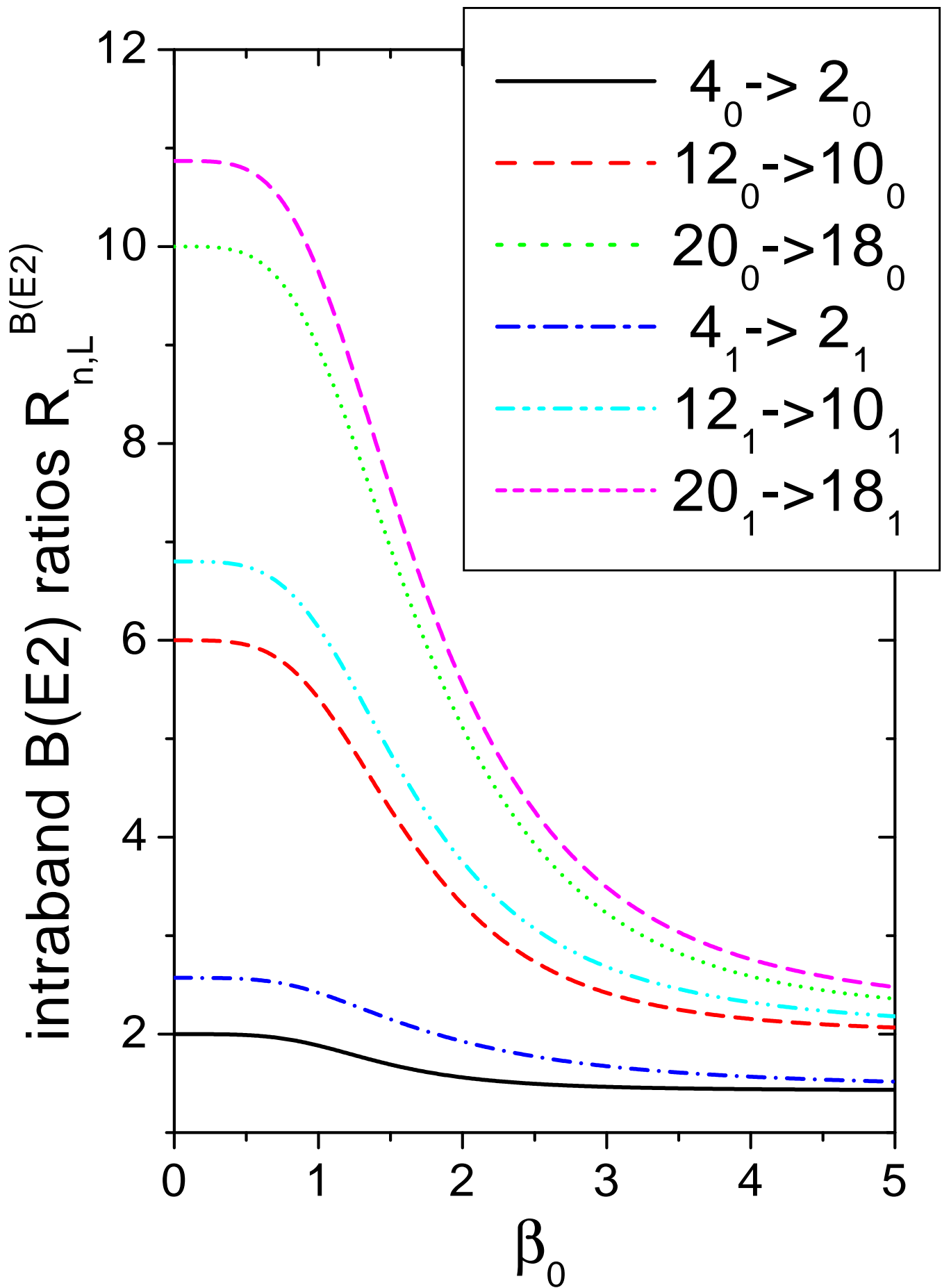
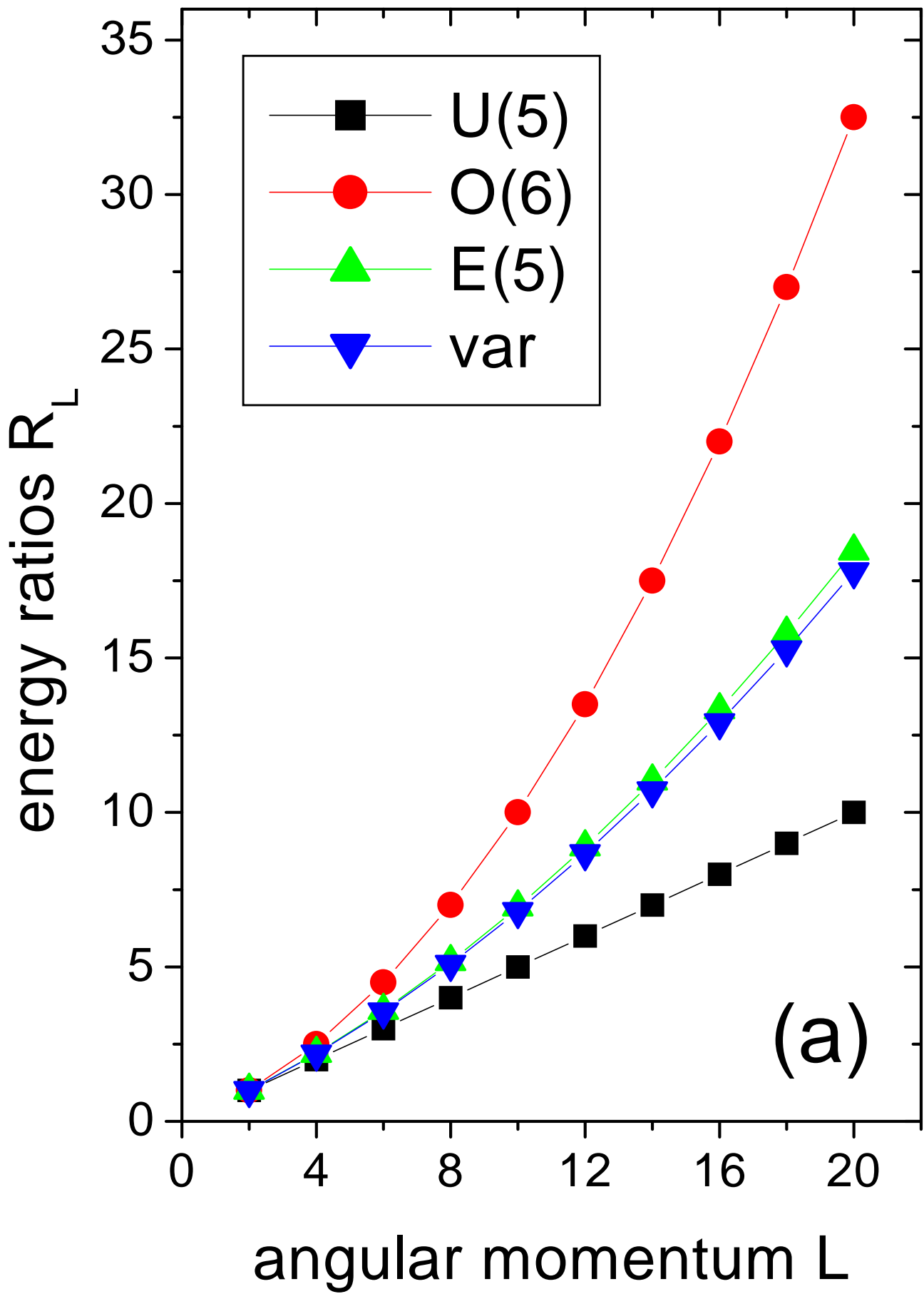
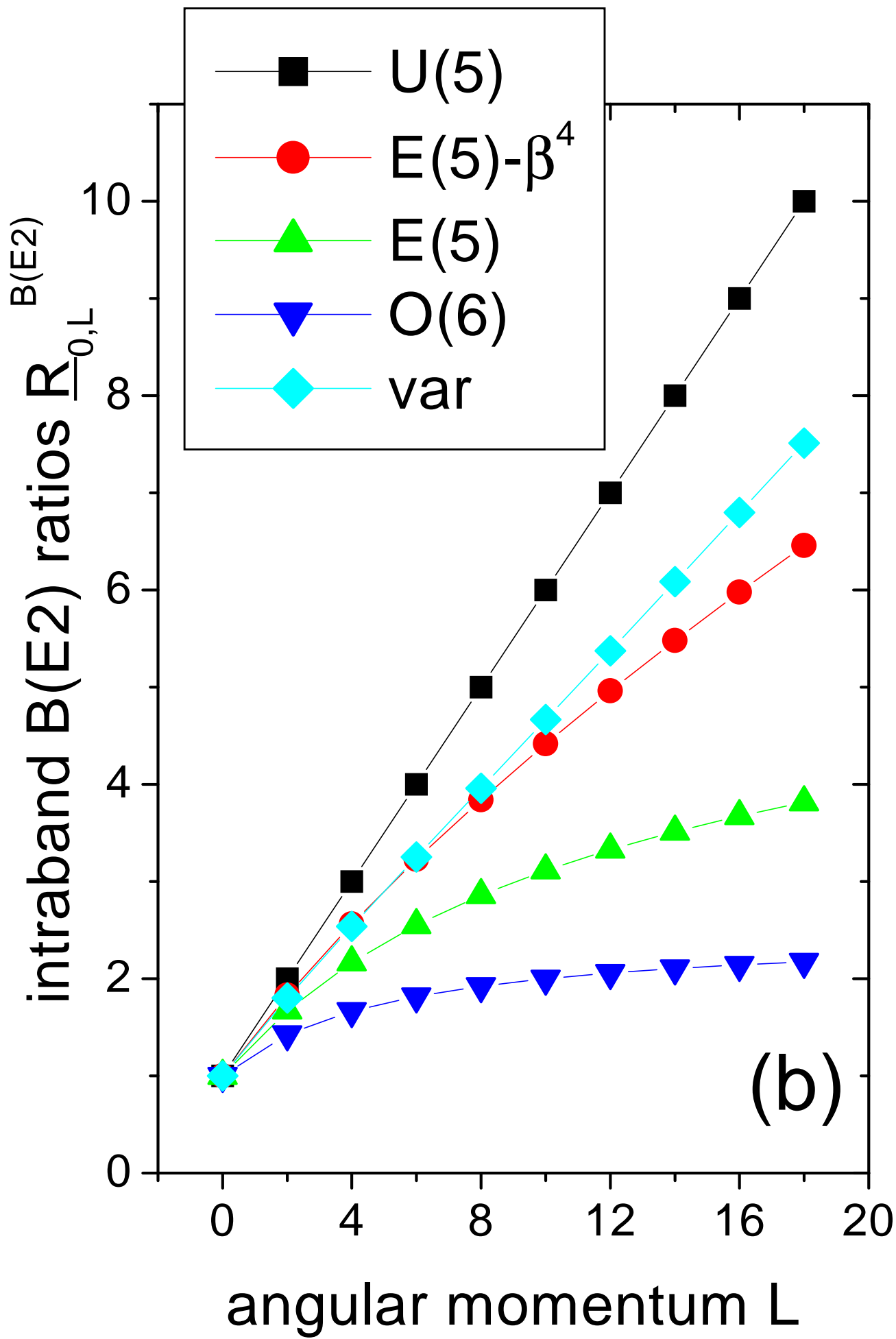
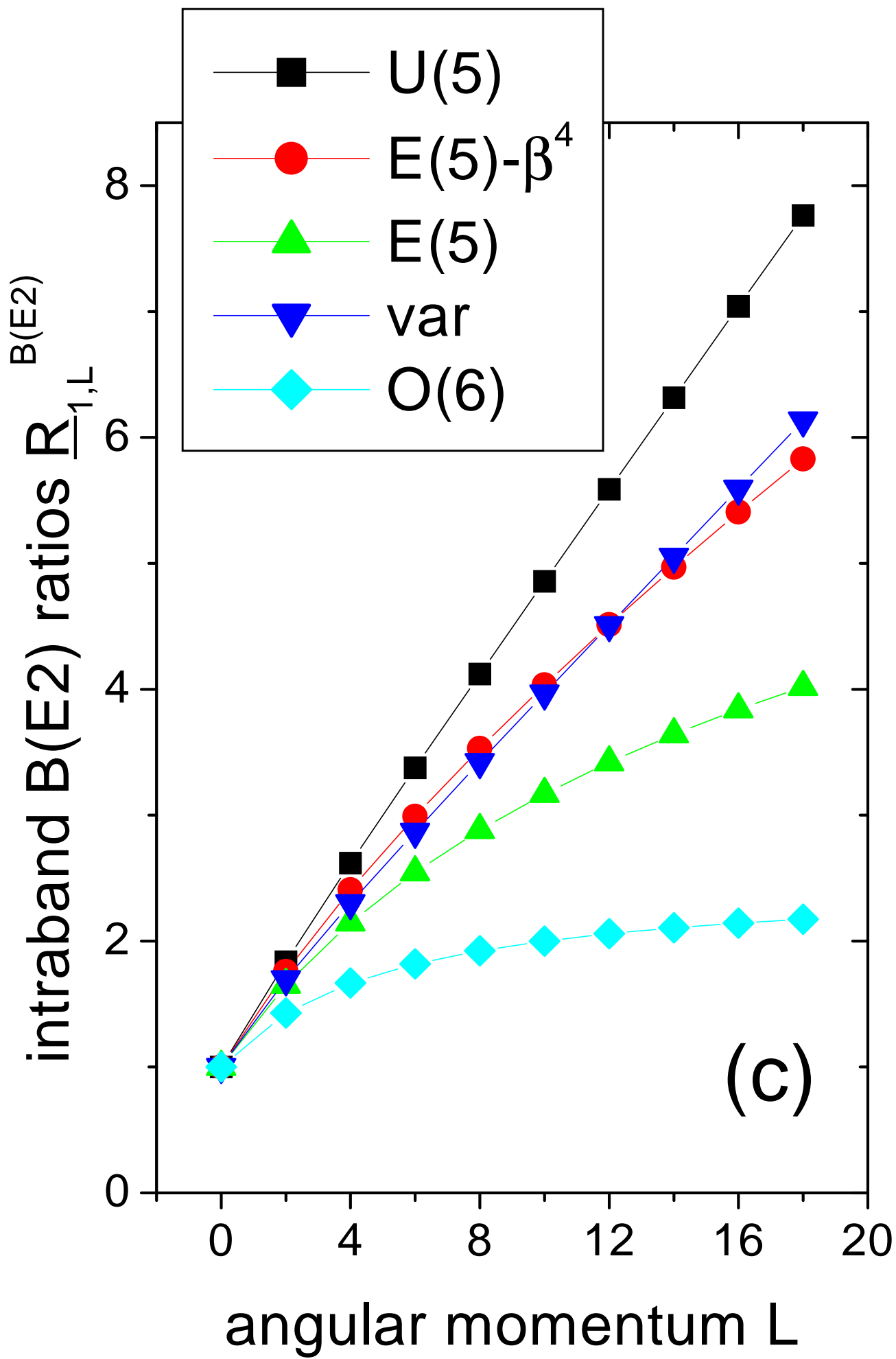
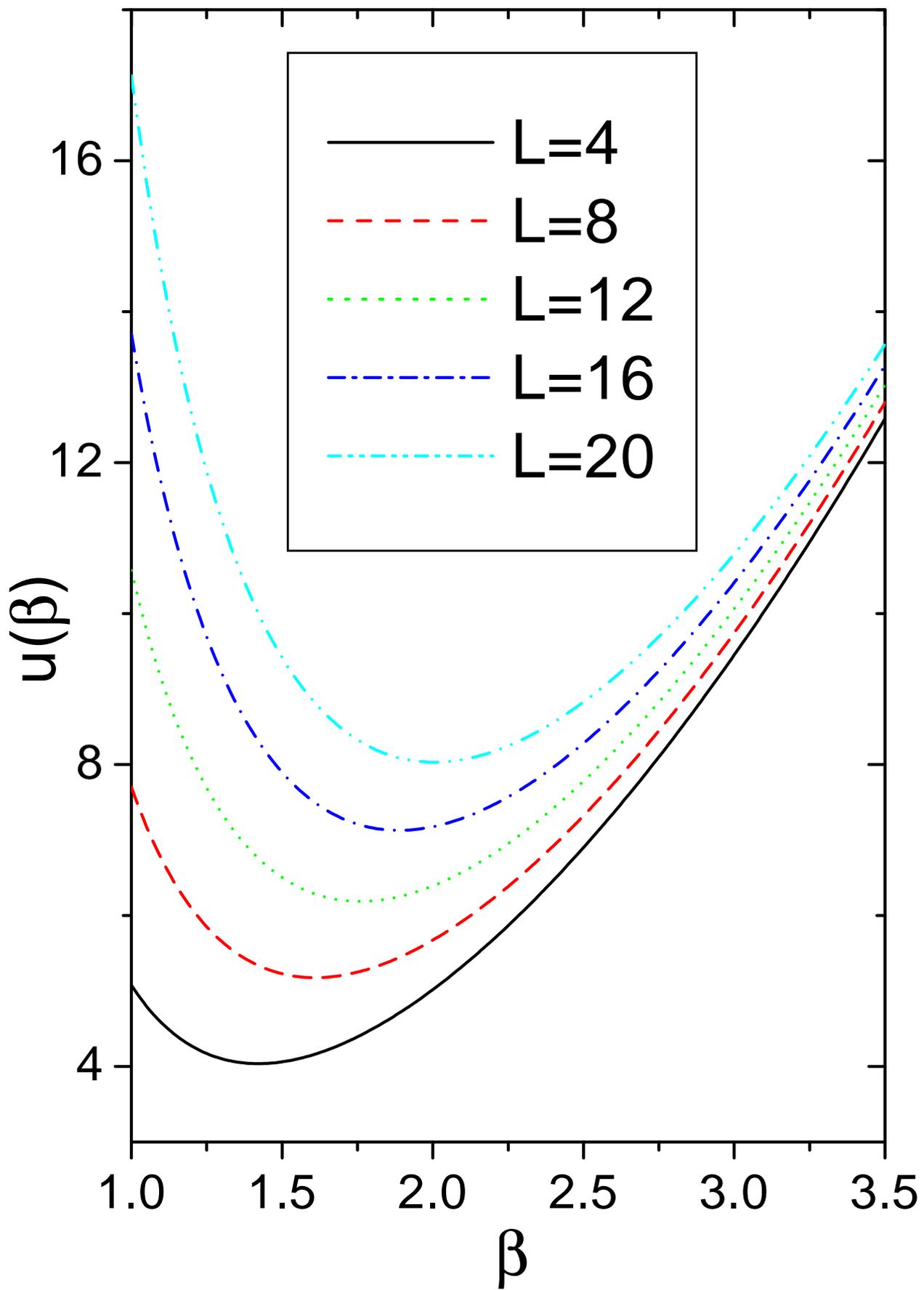
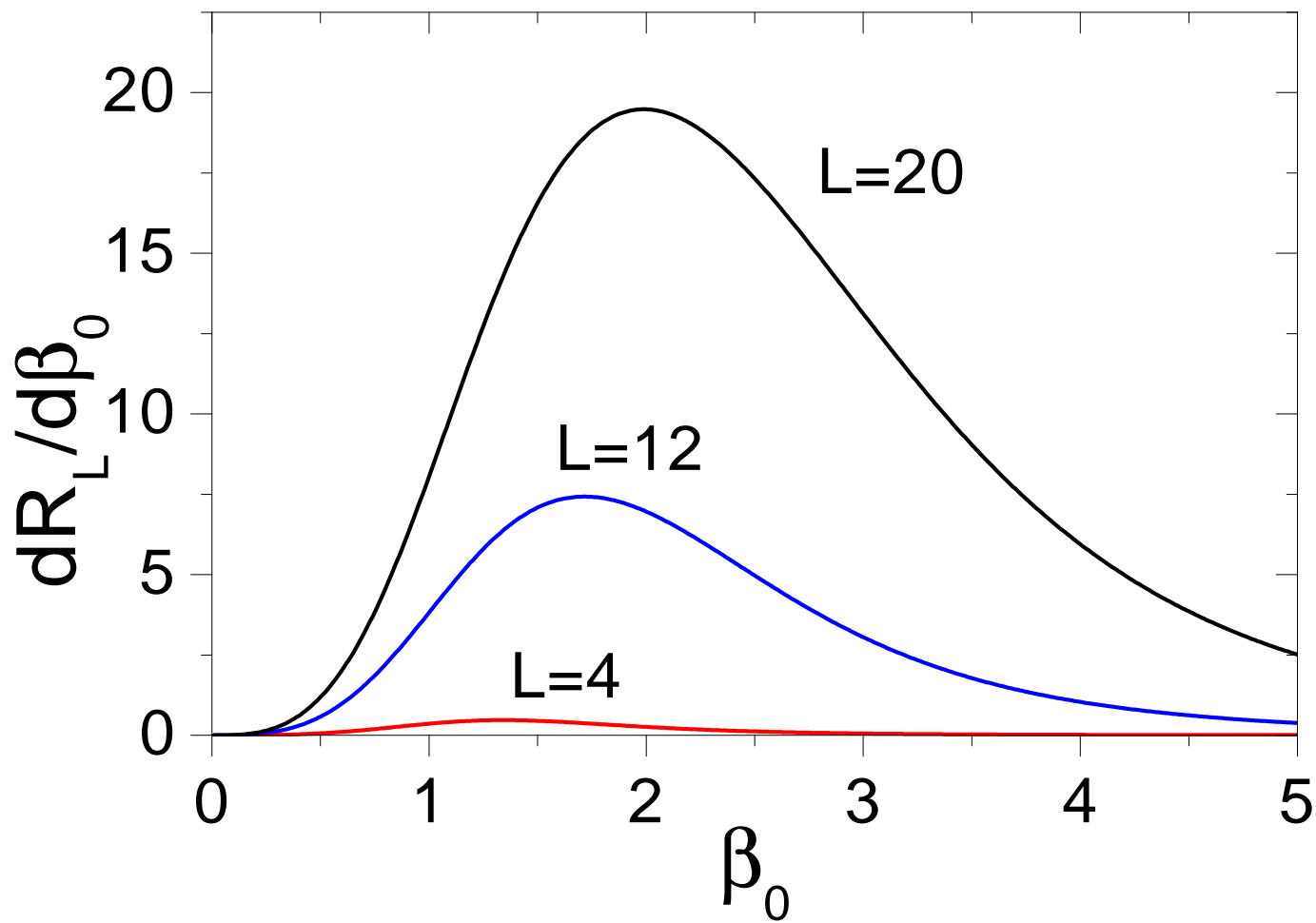
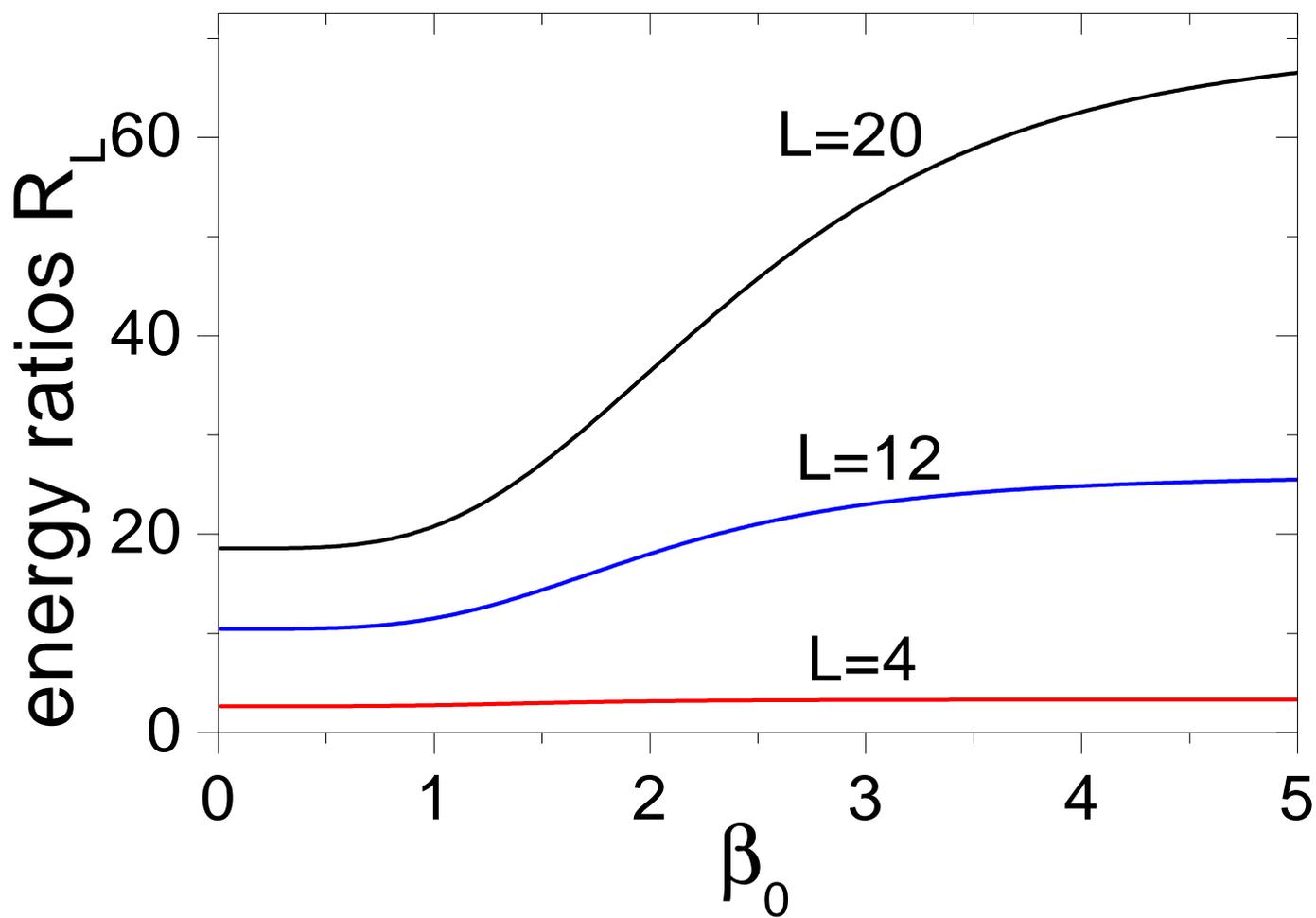
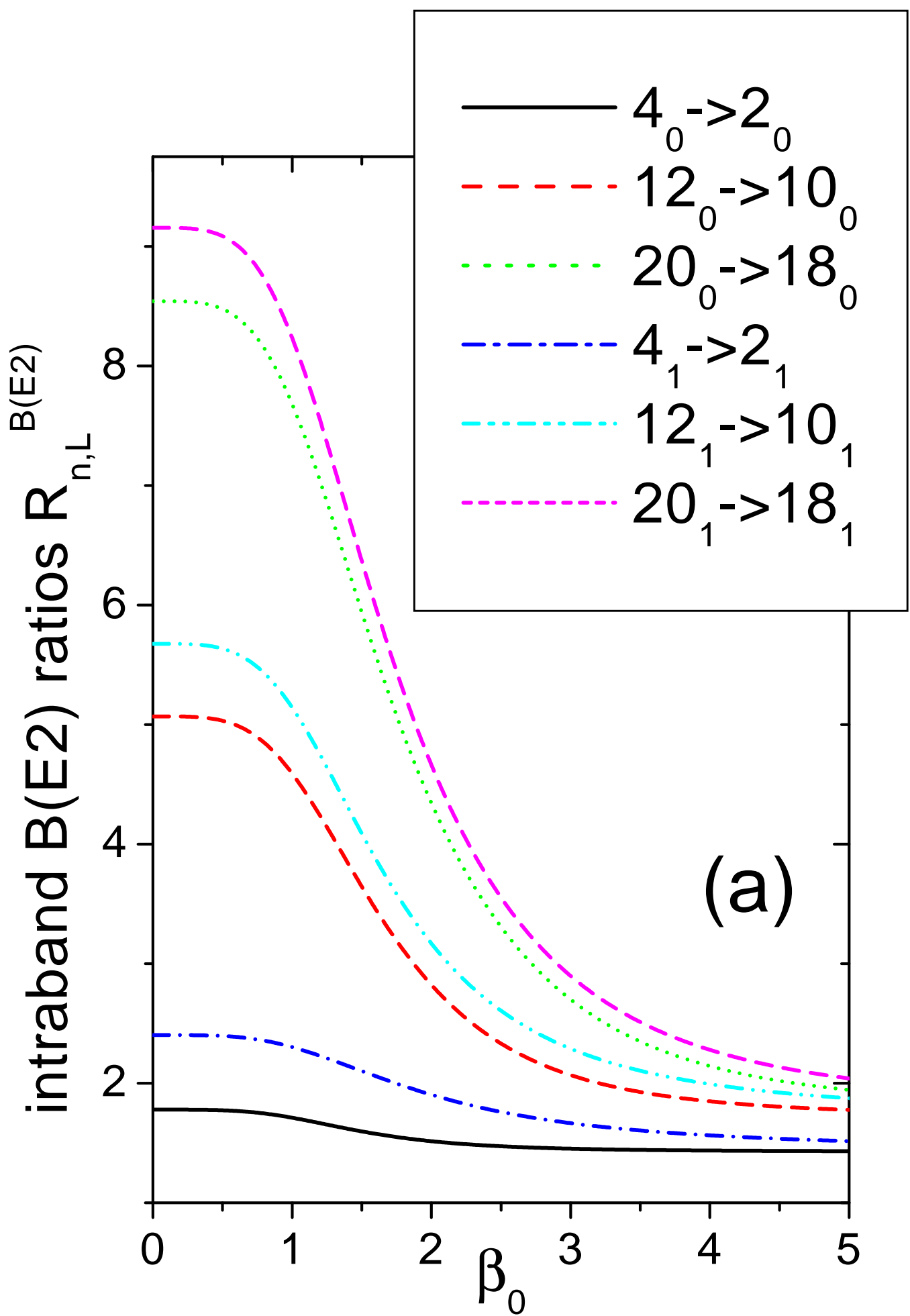
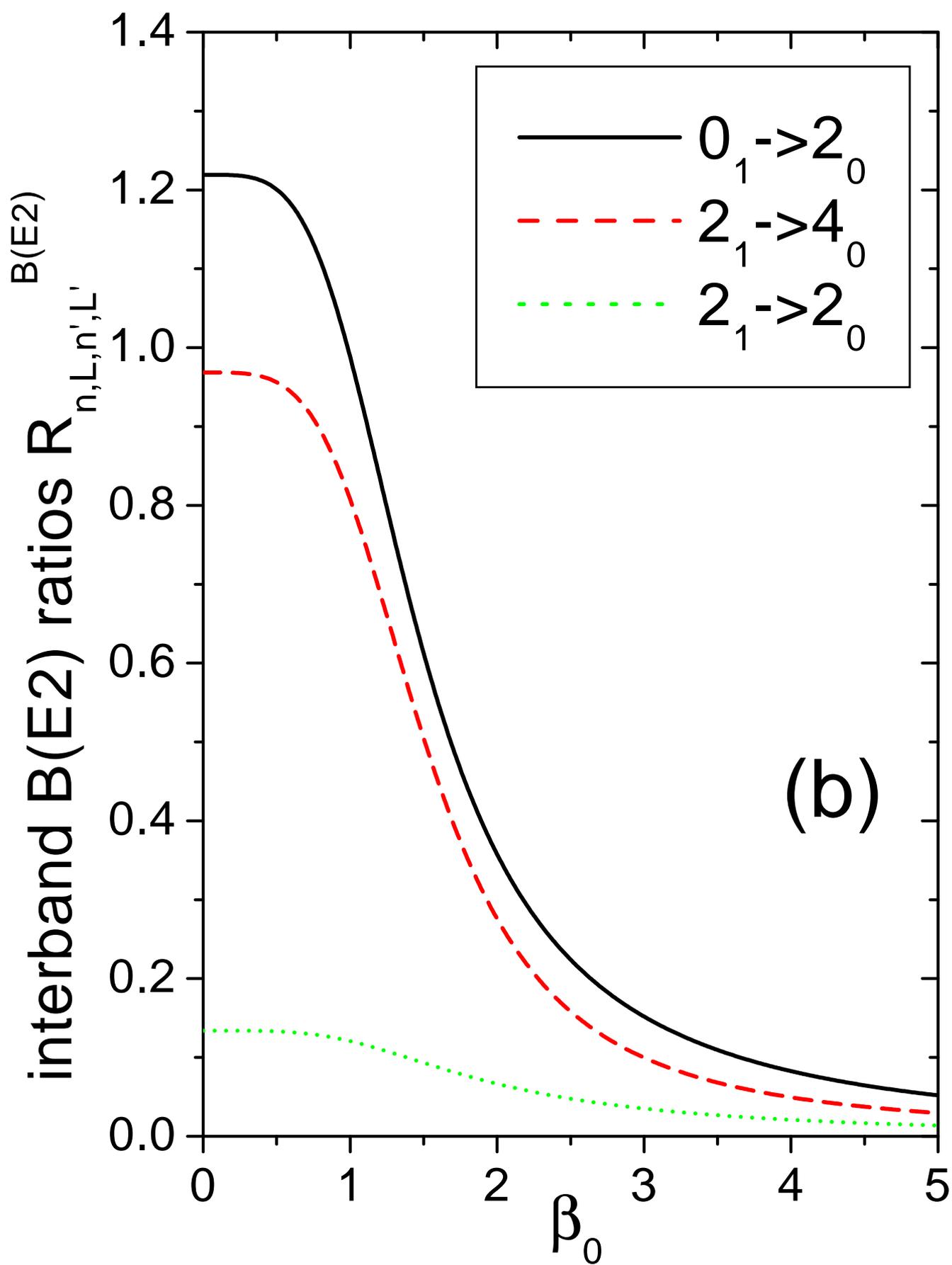
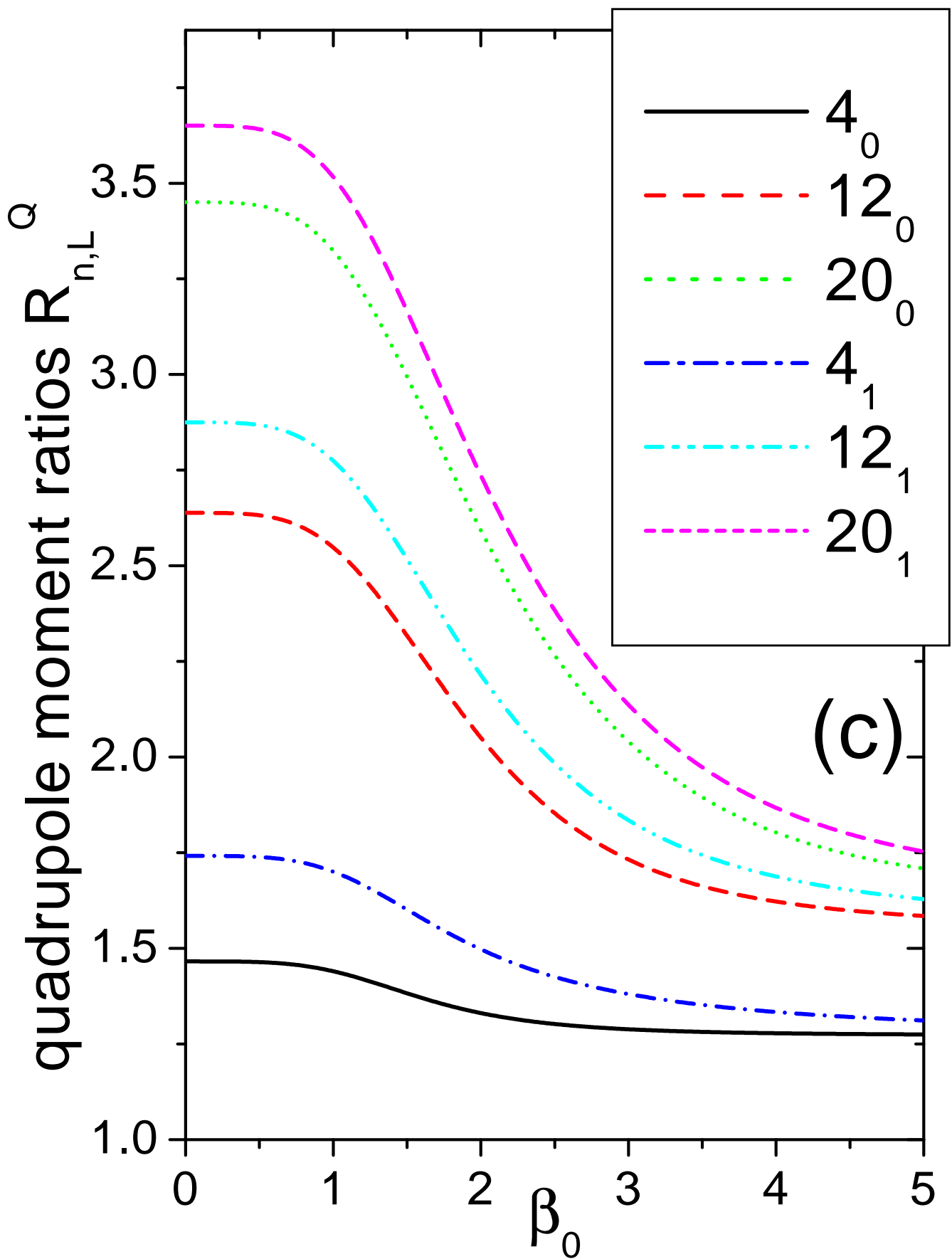
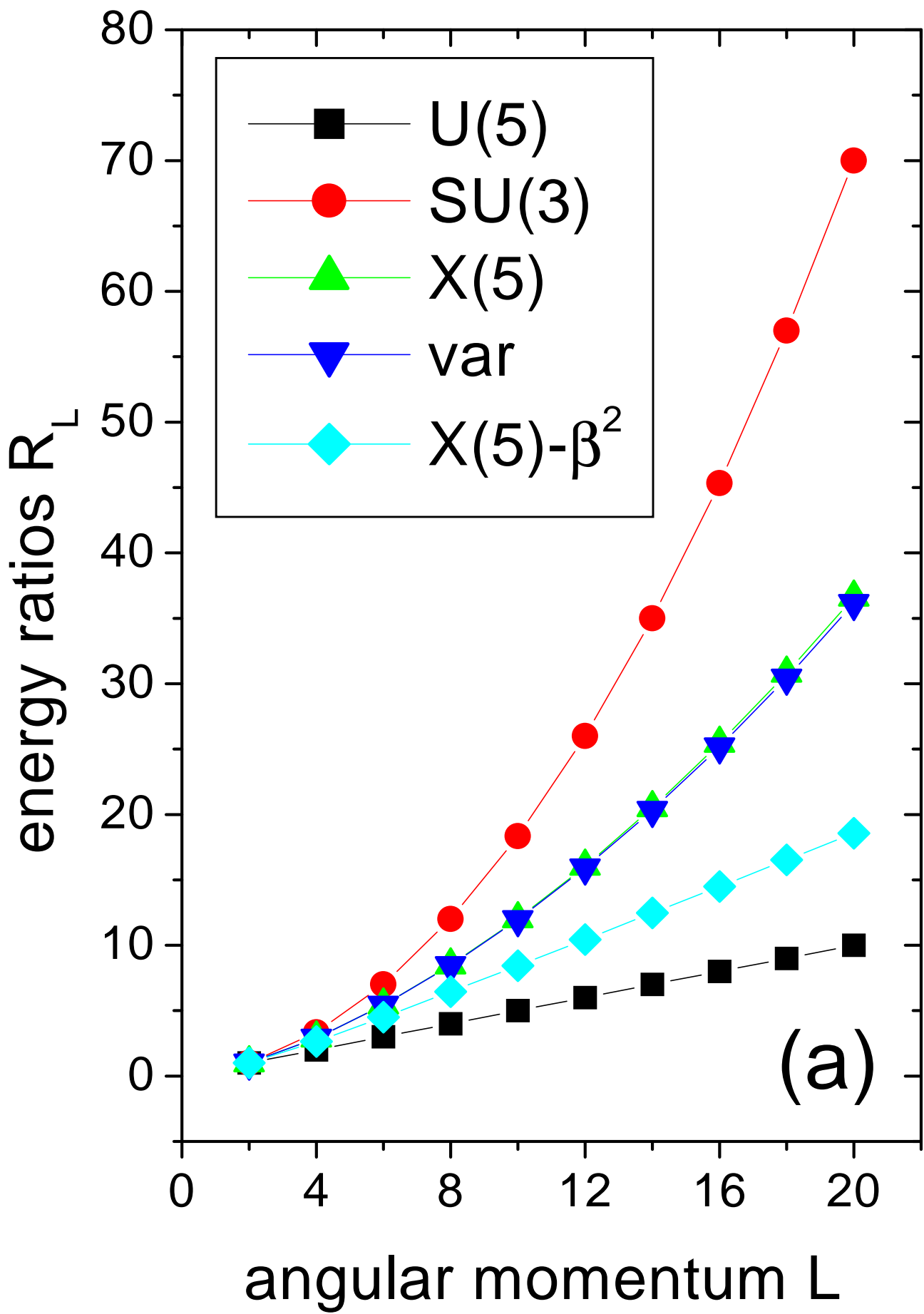
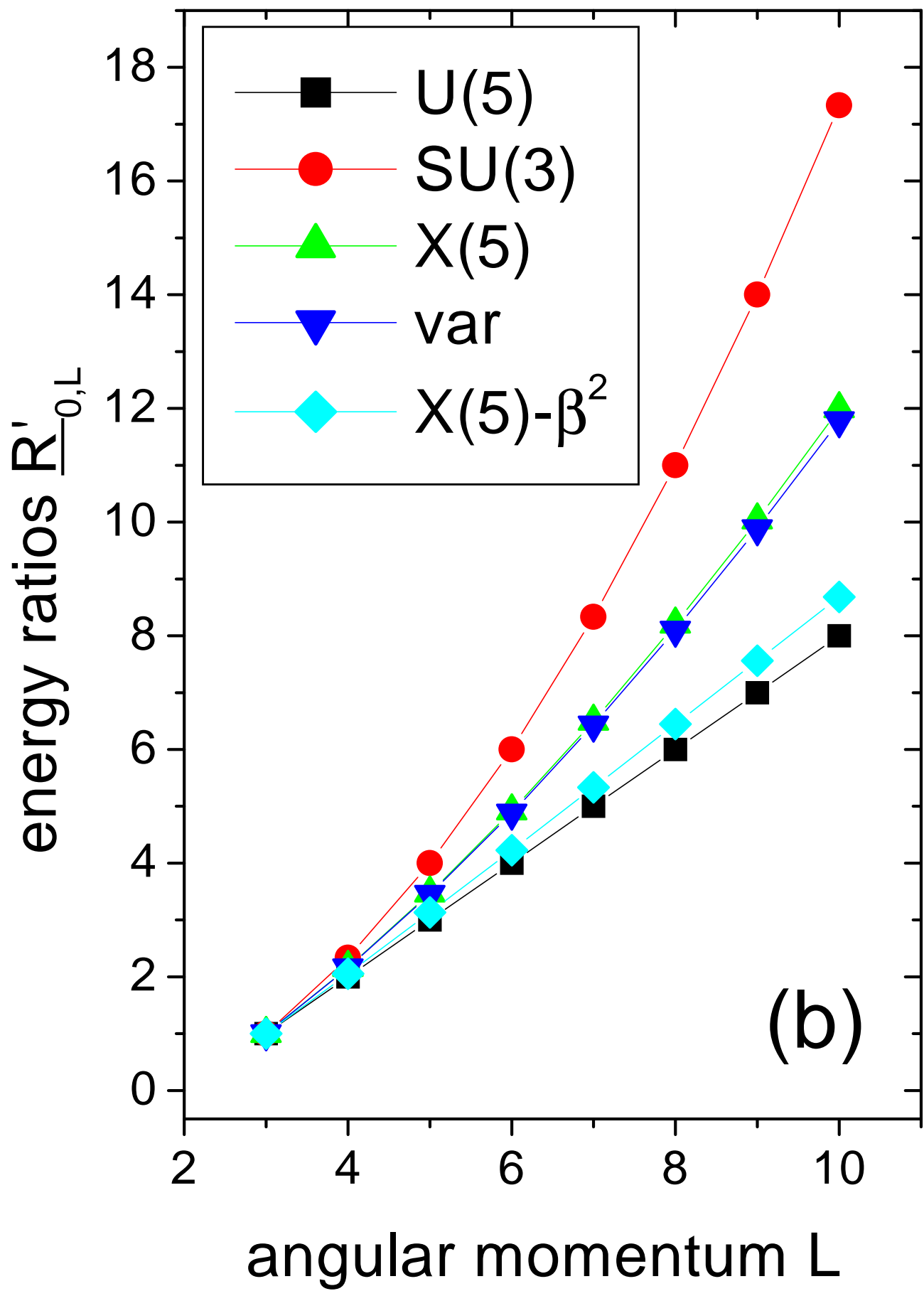
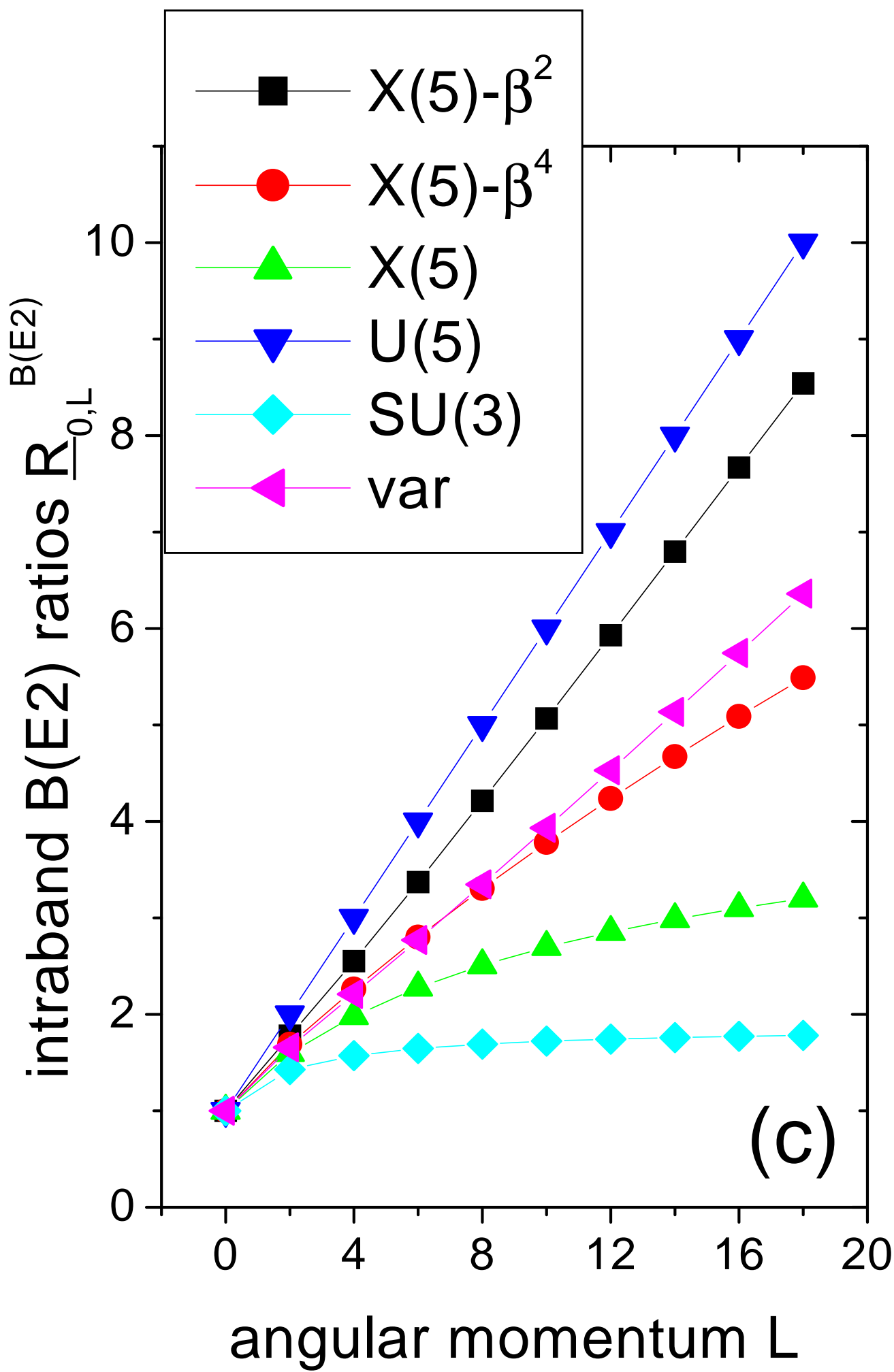
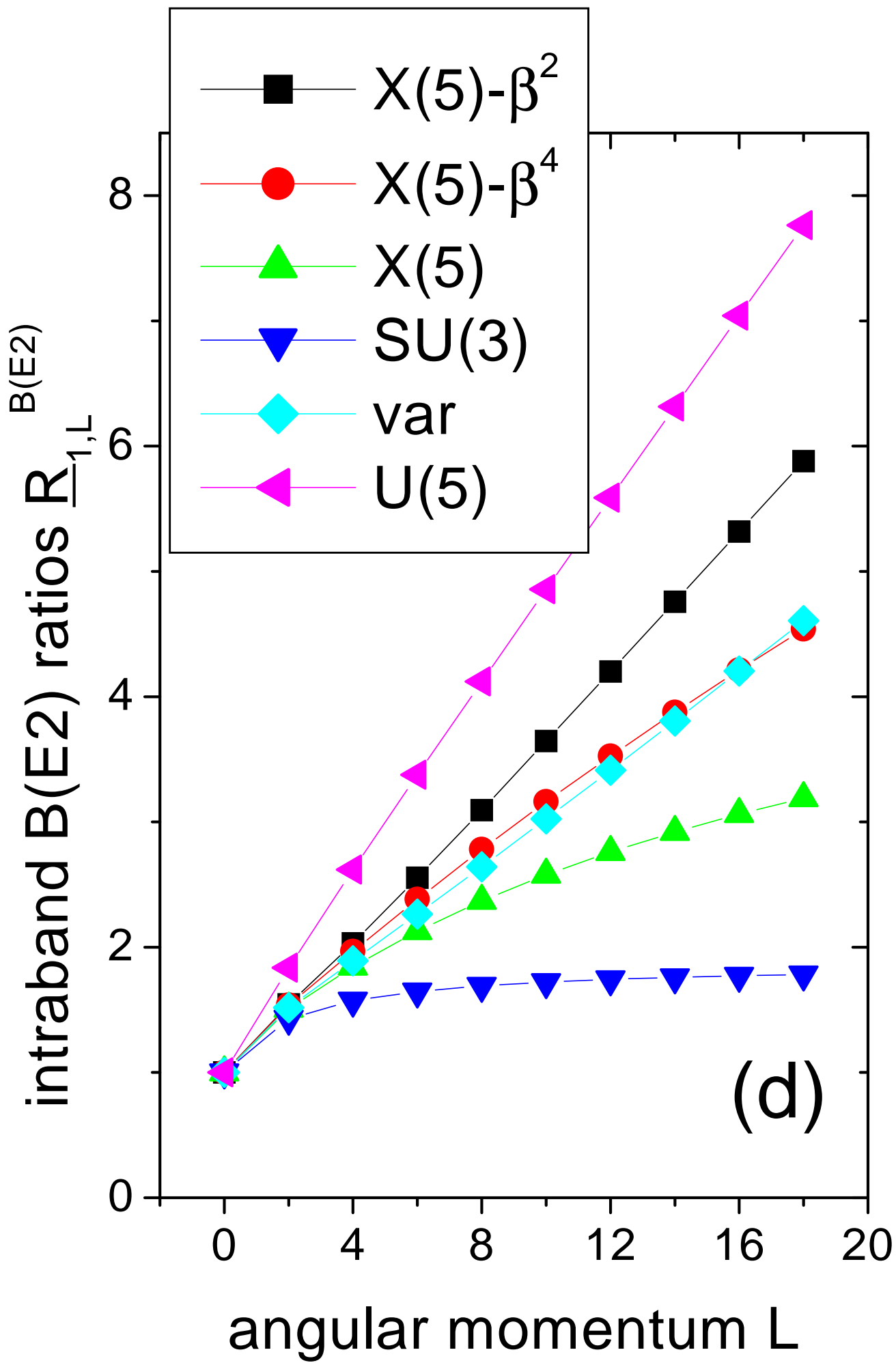
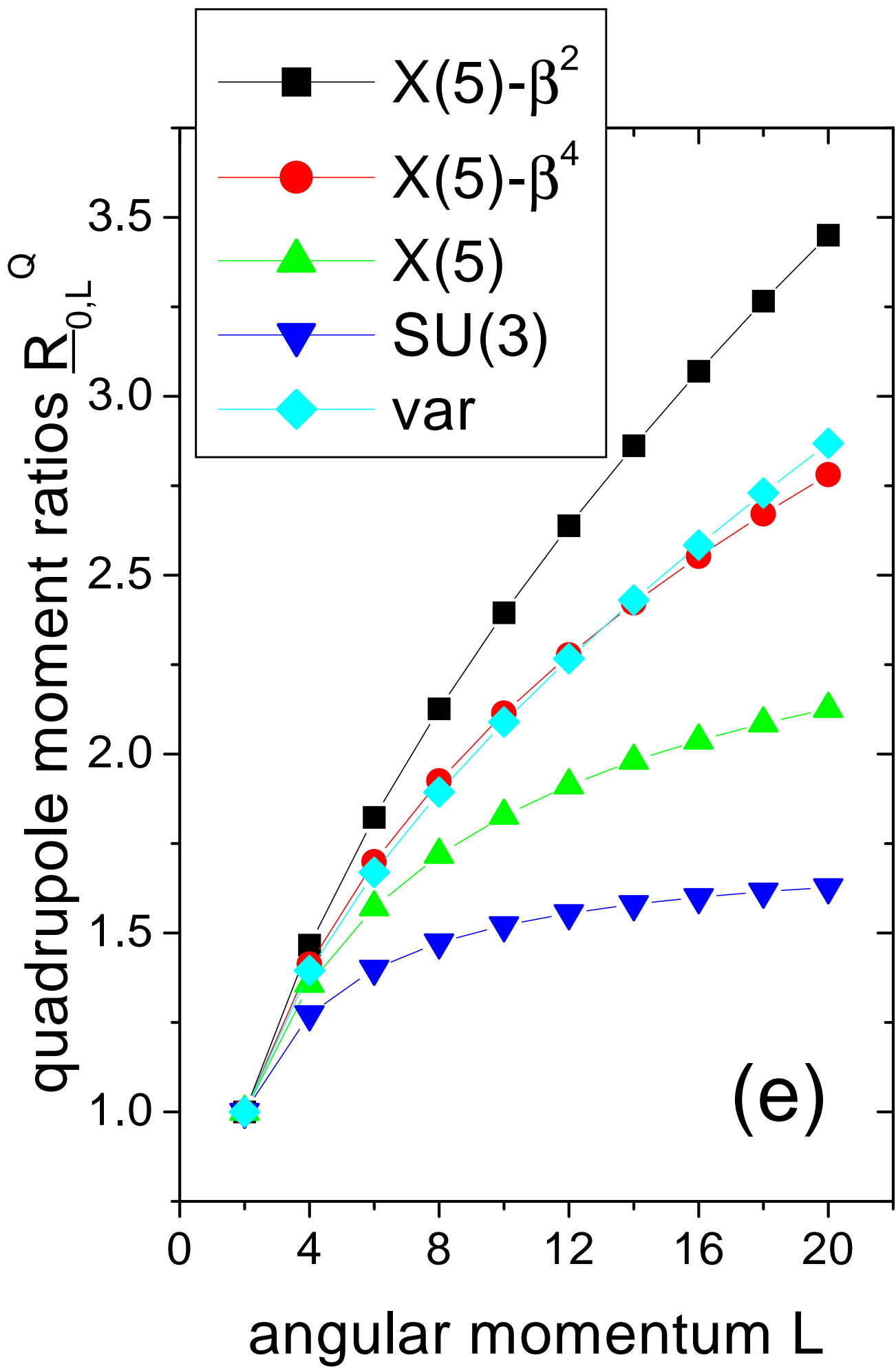
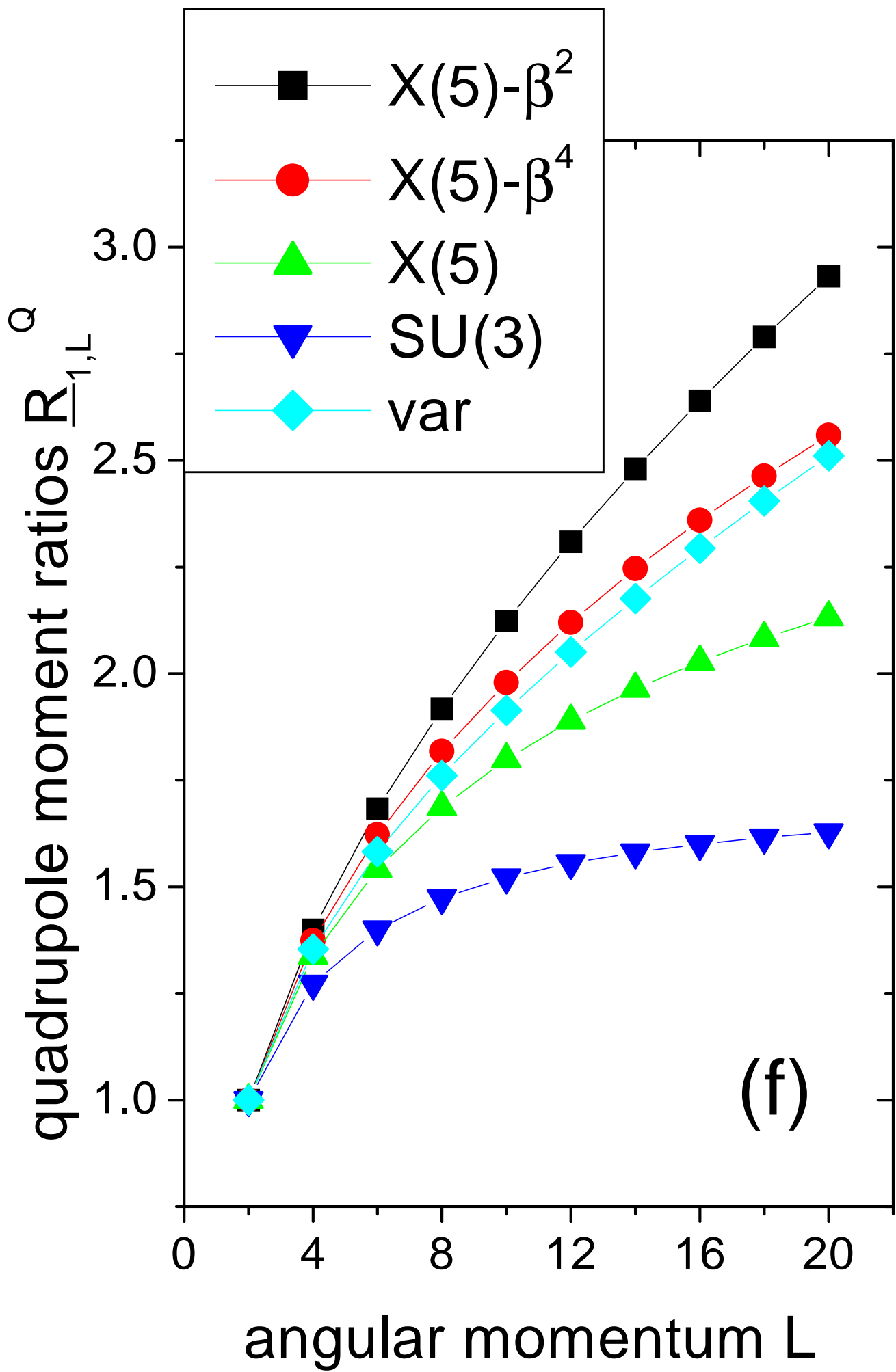
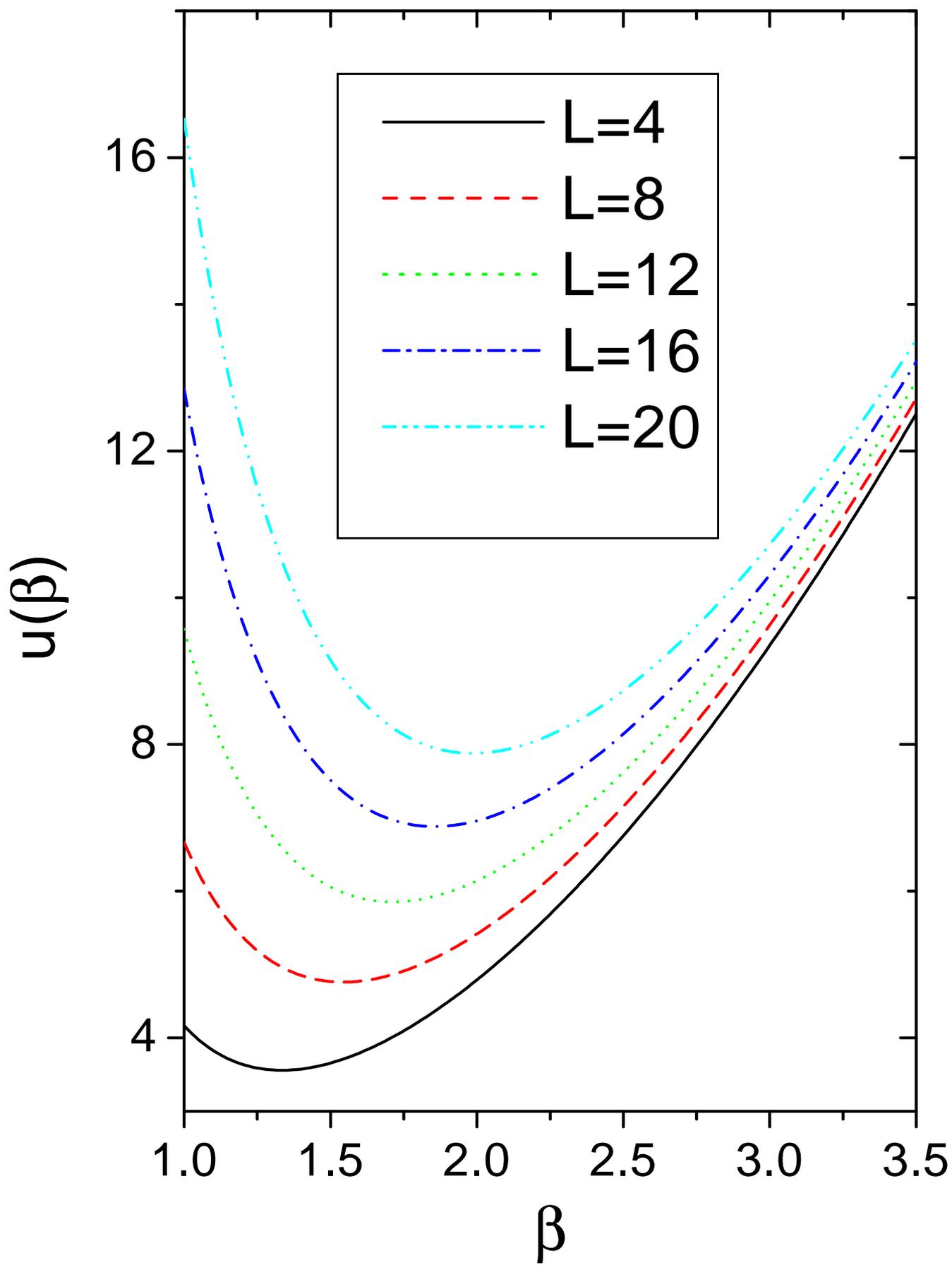
\end{document}